\documentclass[twocolumn,preprintnumbers,amsmath,amssymb,nofootinbib,prd]{revtex4-2}

\usepackage{graphicx}
\usepackage{float}
\usepackage{amsmath}
\usepackage{amssymb}
\usepackage{bm}
\usepackage{cancel}
\usepackage{hyperref}
\usepackage{xcolor}
\usepackage{tikz}
\usepackage[compat=1.1.0]{tikz-feynman}
\usepackage{subfig}

\makeatletter
\long\def\@makecaption#1#2{%
  \vskip\abovecaptionskip
  \sbox\@tempboxa{#1: #2}%
  \ifdim \wd\@tempboxa >\hsize
    {\setlength{\parindent}{0pt}#1: #2\par}%
  \else
    \global\@minipagefalse
    \hb@xt@\hsize{\hfil\box\@tempboxa\hfil}%
  \fi
  \vskip\belowcaptionskip}
\makeatother

\newcommand{\dm}{\chi}

\newcommand{\bea}{\begin{eqnarray}}
\newcommand{\eea}{\end{eqnarray}}

\begin{document}

\title{Dark Matter with a Drag at Low Redshift}

 \author{Martin Schmaltz}
  \thanks{After July 15, 2026: Max Planck Institute for Physics, 85748 Garching, Germany}
  \author{Eashwar N. Sivarajan}
  \affiliation{Physics Department, Boston University, Boston, MA 02215, USA \\
    E-mail: {\tt schmaltz@bu.edu, eashwar@bu.edu}}
 
\begin{abstract}
Recent analyses of $f\sigma_8$ and weak-lensing data indicate that the linear growth rate at $z\lesssim 1$ may be lower than predicted by $\Lambda$CDM. This motivates models of dark matter in which large scale structure growth slows relative to $\Lambda$CDM at late times. We construct particle models in which dark matter experiences a drag with dark radiation that grows at late times, unlike conventional DM--DR interactions, which fade as the universe expands. A key ingredient is that the radiation interacting with the dark matter is produced at late times from dark matter decay. An explicit model, interacting Decaying Cold Dark Matter (iDCDM), adds two parameters beyond $\Lambda$CDM while leaving the background, BBN, and primary CMB intact. But it predicts a step-shaped suppression of the linear growth rate $f(k,z)$, a distinctive target for DESI, Euclid, and Rubin. Confronted with current data, iDCDM shows a modest preference over $\Lambda$CDM, driven by $f\sigma_8$, with $\Delta\chi^2$ between $-2.7$ and $-7.6$ depending on the assumed scaling of the drag with redshift and on neutrino masses. The decisive test will come from upcoming $k$- and $z$-resolved growth measurements.
\end{abstract}

\maketitle

\section{Introduction}
\label{sec:intro}

Recent analyses of $f\sigma_8$ and weak-lensing data indicate that the linear growth rate at $z\lesssim 1$ may be lower than predicted by $\Lambda$CDM~\cite{Macaulay:2013swa,Huterer1,Terasawa:2025fpf,Lin:2023uux,Adil:2023jtu,Escamilla:2025imi} (but see also~\cite{Lai:2025xkf,Qin:2025ggt}).
This points to physics beyond $\Lambda$CDM whose effect on perturbations turns on at late times.
Modifications of gravity~\cite{Koyama2016,Nesseris:2017vor,Kazantzidis:2019nuh,Skara:2019usd} offer one route, but constructing UV-complete realizations consistent with solar-system and laboratory tests is difficult.
We instead explore particle-physics alternatives.

The next generation of galaxy surveys, DESI~\cite{DESIDR2,DESIfsig8}, Euclid, and the Vera C.\ Rubin Observatory, will measure the matter power spectrum $P(k,z)$ from $z = 0$ to $z\sim 2$ at percent-level precision.
Physically motivated models whose predictions for $P(k,z)$ differ from $\Lambda$CDM in their $k$- and $z$-dependence are therefore interesting both as candidate explanations of the current hints and as concrete targets for upcoming data.

A natural particle model that modifies growth is a dark matter component that decays into radiation at late times.
However, this does not by itself modify the matter density contrast $\delta_m$: the decay rate is independent of the local environment, so an overdense region loses the same fraction of its matter as one at mean density, leaving $\delta_m = \delta\rho_m/\bar\rho_m$ unchanged to leading order.
Subleading effects exist (shallower potentials from depleted matter, modified background expansion), but also modify the background; CMB and BAO data tightly constrain those backgrounds~\cite{Enqvist2015,Poulin1,Poulin2}, so decay alone does not significantly suppress $P(k)$.

Elastic scattering between dark matter (DM) and a dark-radiation (DR) component, by contrast, acts directly on the perturbations~\cite{ManuelNADM}.
The scattering introduces a drag force on the matter: on sub-horizon scales, dark matter falls into gravitational potential wells while the DR, supported by its own pressure, does not follow, but instead streams out of overdense regions. Elastic scattering between the two fluids in relative motion sources a momentum exchange that slows gravitational infall of the DM and suppresses the growth of matter density perturbations.
The momentum-transfer rate, or drag rate, depends on the $k$-dependent relative bulk velocity between the fluids: small-scale modes are suppressed, super-horizon modes are unaffected.

This idea has been developed extensively into a phenomenological program in which a primordial DR component is present since the early universe~\cite{ManuelNADM,Lesgourgues:2015wza,Buen-Abad:2017gxg,Aloni:2021eaq,Garny:2025kqj,Buen-Abad:2025bgd}, recently confronted with full-shape galaxy clustering~\cite{Rubira:2022xhb}, cluster abundances~\cite{SPT:2024roj}, ACT DR6~\cite{Cvetko:2025kda}, and the Lyman-$\alpha$ forest~\cite{Bagherian:2024obh}.
In all of these, the DR energy density dilutes as $a^{-4}$ and the ratio $\Gamma_s/H$ that controls the drag drops with time.
The DM--DR friction is therefore most active in the early universe and is negligible by the redshifts $z\lesssim 1$ where late-time growth measurements are most precise.
This is the opposite of what the data motivation above calls for.

What is needed is a drag that \emph{grows} at late times.
This can be achieved in two physically distinct ways: by letting the DR energy density grow relative to matter, or by introducing a drag rate that decreases more slowly than Hubble as the DR cools and dilutes.
The second route was recently considered in~\cite{Dallari:2026wvj}, which studies a phenomenological drag rate $\Gamma_s\propto T_{d}$ in the ETHOS~\cite{Cyr-Racine:2015ihg,Vogelsberger:2015gpr} framework with a DR bath of conventional $T_d^4 \propto a^{-4}$ dilution: as the DR cools, $\Gamma_s/H$ grows during matter domination. Here we explore the first route with a UV-natural particle model in which slowly decaying dark matter continuously replenishes the DR bath, driving its density to an attractor that dilutes only as $T_d^4 \propto a^{-3/2}$. Then even with a standard scaling of the drag rate, $\Gamma_s\propto T_d^2$, the slow DR dilution keeps $\Gamma_s/H$ growing and the drag active at low redshift.

We dub the resulting model interacting Decaying Cold Dark Matter (iDCDM).%
\footnote{Aspects of this work were presented at the 2025 ``No stone unturned'' University of Utah workshop (slides at \texttt{[https://www.physics.utah.edu/no-stone-unturned/]}).}
Dark matter decays into dark radiation with rate $\Gamma_d$ and elastically scatters off the radiation it produces with momentum-transfer (drag) rate $\Gamma_s$.
Continuous replenishment drives the DR density to an attractor $\rho_\mathrm{DR}\propto \rho_\mathrm{DM}\, \Gamma_d/H\propto a^{-3/2}$, which dilutes slower than matter.
iDCDM adds two free parameters beyond $\Lambda$CDM (the decay and drag rates), leaves the background expansion, BBN, and the primary CMB intact, and predicts a step-shaped suppression of the linear growth rate $f(k,z)$ as a distinctive target for the upcoming surveys.

The model can also bear on the $S_8$ tension~\cite{Abdalla2022,DESY3,KiDS1000,DESIS8} (the KiDS-Legacy analysis finds $S_8$ consistent with Planck~\cite{KiDSLegacy}): the DM--DR interaction suppresses $P(k)$ at late times, so low-redshift probes find lower $S_8$ than high-redshift determinations from the CMB and CMB lensing.

The paper is organized as follows.
In Section~\ref{sec:model} we present the UV model and reduce it to an effective cosmological description that extends $\Lambda$CDM by the decay rate $\Gamma_d$, the drag rate $\Gamma_s^0$ today, and a discrete index $n_{\Gamma_s}$ parametrizing the temperature scaling of $\Gamma_s$, together with the background and perturbation equations.
Section~\ref{sec:effects} discusses the observable signatures: the step-shaped suppression of $P_m(k,z)$, the scale-dependent growth rate, and the model-dependence of $f\sigma_8$ measurements.
Section~\ref{sec:data} describes the datasets and sampling.
Section~\ref{sec:results} reports parameter constraints and the impact of neutrino masses.
We conclude in Sec.~\ref{sec:conclusions}.

\section{The Model}
\label{sec:model}

\subsection{UV model}
\label{sec:uv}

One possible particle physics model of iDCDM is an unbroken dark $U(1)$ gauge theory,
with a small dark gauge coupling $\alpha_d = g_d^2/(4\pi)$ and three particles: a massive charged fermion $\chi$ which is the dark matter, a massless charged fermion $\psi$ and a massless dark photon $\gamma_d$ which together form the dark radiation.

At the renormalizable level, the dark matter is stable by gauge invariance, but dark matter decays
\begin{equation}
\chi \to \psi + \gamma_d
\end{equation}
are mediated by a dimension-5 magnetic dipole operator with a rate $\Gamma_d$ which can naturally be very small. We will treat $\Gamma_d$ as a free parameter, but for the physics that we are interested in, $\Gamma_d/H_0 \lesssim 10^{-2}$, with even smaller values phenomenologically preferred.
The three-body decay $\chi \to \psi + \psi + \bar\psi$ is suppressed by an additional factor of $\alpha_d/(4\pi)$ and therefore negligible.

The light decay products $\psi$ and $\gamma_d$ thermalize rapidly through gauge interactions at a rate $\Gamma_\mathrm{th}\sim\alpha_d^2\,T_d$. This is always much larger than the Hubble rate in the parameter space of interest as will become clear in the next paragraph (see also App.~\ref{app:uv}).
They form a relativistic fluid with temperature $T_d$, energy density $\rho_\mathrm{DR} = \frac{11}{2} \frac{\pi^2}{30} T_d^4$, and no anisotropic stress.

The leading momentum transfer between the dark matter and dark radiation fluids proceeds via $\chi\,\psi \to \chi\,\psi$ scattering with dark photon exchange. The momentum transfer rate is suppressed relative to the dark radiation thermalization rate by the small dimensionless ratio $T_d/M_\chi \ll 1$~\cite{ManuelNADM}:
\begin{equation}
\label{eq:Gammas_UV}
\Gamma_s \sim \alpha_d^2\,\frac{T_d^2}{M_\chi}\,.
\end{equation}
We will be interested in this rate being $\sim 10^{-2} H_0$ today. 
This implies that $\alpha_d\ll 1$ and $\Gamma_\mathrm{th}/H_0 \gg 1$ as long as $T_d\ll M_\chi\ll M_{\rm Pl}$. As a benchmark, taking $\alpha_d = 10^{-8}$, $M_\chi = 65$~GeV and a dark radiation temperature $T_{d,0}= 10^{-4}$~eV gives $\Gamma_s^0/H_0= 1\%$.

To eliminate unobservable parameters, we write the $T_d^2$ dependence as $\Gamma_s \propto \rho_\mathrm{DR}^{1/2}$ and parameterize the scattering rate as
\begin{equation}
\label{eq:Gammas_param}
\Gamma_s(a) = \Gamma_s^0\left(\frac{\rho_\mathrm{DR}(a)}{\rho_\mathrm{DR}^0}\right)^{n_{\Gamma_s}/4},
\end{equation}
with $\Gamma_s^0 \equiv \Gamma_s(a=1)$.
The particle physics model described above predicts $n_{\Gamma_s}=2$, but other microphysics models give different temperature dependences of the DM-DR scattering cross section: for example, Compton scattering predicts $n_{\Gamma_s}=4$ while pure scalar models with trilinear scalar couplings give $n_{\Gamma_s}=0$. Alternatively, a different time evolution of the radiation energy density would also lead to a different effective $n_{\Gamma_s}$. We give the derivation of $\Gamma_s$ and discuss alternative models for $n_{\Gamma_s}=0$ in App.~\ref{app:uv}.

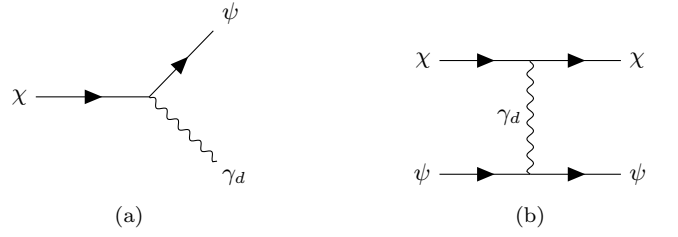
\begin{figure}[t]
\centering
\subfloat[]{%
\begin{tikzpicture}
\begin{feynman}
\vertex (a) [dot];
\vertex [above right=1.2cm of a] (f1) {\(\psi\)};
\vertex [below right=1.2cm of a] (f2) {\(\gamma_d\)};
\vertex [left=1.5cm of a] (i1) {\(\dm\)};
\diagram* {
(i1) -- [fermion] (a) -- [fermion] (f1),
(a) -- [boson] (f2),
};
\end{feynman}
\end{tikzpicture}%
}%
\hspace{2cm}%
\subfloat[]{%
\begin{tikzpicture}
\begin{feynman}
\vertex (a);
\vertex [below=1.5cm of a] (b);
\vertex [left=1.2cm of a] (i1) {\(\dm\)};
\vertex [right=1.2cm of a] (f1) {\(\dm\)};
\vertex [left=1.2cm of b] (i2) {\(\psi\)};
\vertex [right=1.2cm of b] (f2) {\(\psi\)};
\diagram* {
(i1) -- [fermion] (a) -- [fermion] (f1),
(i2) -- [fermion] (b) -- [fermion] (f2),
(a) -- [boson, edge label'=\(\gamma_d\)] (b),
};
\end{feynman}
\end{tikzpicture}%
}
\caption{(a) Dark matter decay $\dm \to \psi + \gamma_d$. (b) Dark photon-mediated DM--DR scattering $\dm\,\psi \to \dm\,\psi$.}
\label{fig:feynman}
\end{figure}

The two processes, decay and scattering, are shown in Fig.~\ref{fig:feynman}.

\subsection{Summary of effective cosmological description}
\label{sec:effective}

At the cosmological level, the UV model reduces to a two-component dark sector with DM and DR.
The DM $\chi$ is a cold, pressureless fluid that decays at rate $\Gamma_d$ and elastically scatters off the DR leading to momentum-transfer between the two fluids at the rate $\Gamma_s(a)$.
We assume that the dark radiation (DR) energy density is initially negligible and is completely specified by the DM decay rate at late times. Since it has frequent self-interactions, the radiation perturbations are those of a relativistic perfect fluid ($w = 1/3$, $\sigma_\mathrm{DR} = 0$) with source terms proportional to $\Gamma_d$ from the decay. In addition the DR velocity perturbations are coupled to the DM velocity perturbations through the momentum exchange from the DM--DR scattering. The temperature (and therefore scale-factor) dependence of this momentum exchange is UV-model dependent; we parameterize it as $\Gamma_s = \Gamma_s^0 (\rho_\mathrm{DR}/\rho_\mathrm{DR}^0)^{n_{\Gamma_s}/4}$ [Eq.~\eqref{eq:Gammas_param}], where $n_{\Gamma_s}=2$ corresponds to the $T_d^2$ scaling of dark QED. We also study the $n_{\Gamma_s}=0$ case which corresponds to constant $\Gamma_s$. 

Thus the full cosmological model has two continuous and one discrete parameter beyond $\Lambda$CDM
\bea
\Gamma_d \qquad \Gamma_s^0\qquad n_{\Gamma_s}
\eea

\subsection{Background evolution}
\label{sec:background}

The background energy densities of the two dark-sector components obey the coupled continuity equations
\begin{align}
\dot \rho_\chi &= -3\mathcal{H}\rho_\chi - a \Gamma_d\,\rho_\chi\,,\label{eq:bg_chi}\\
\dot \rho_\mathrm{DR} &= -4\mathcal{H}\rho_\mathrm{DR} + a \Gamma_d\,\rho_\chi\,,\label{eq:bg_DR}
\end{align}
where dots denote derivatives with respect to conformal time $\tau$ and $\mathcal{H} = \dot a/a$ is the conformal Hubble rate. 
The first terms on the right-hand side are the standard dilution from expansion: $a^{-3}$ for matter and $a^{-4}$ for radiation.
The remaining terms transfer energy density from the DM to the DR at rate $\Gamma_d$.

Continuous decay replenishment drives the DR density toward an attractor solution. Assuming matter domination and to leading order in $\Gamma_d/H$, this solution is
\begin{equation}
\label{eq:attractor}
\rho_\mathrm{DR} \simeq \frac{2}{5}\,\frac{\Gamma_d}{H}\,\rho_\chi \propto a^{-3/2}\,.
\end{equation}
Primordial radiation dilutes as $a^{-4}$ and becomes negligible relative to matter ($\rho_m\propto a^{-3}$) at late times.
Decay-produced DR, by contrast, dilutes as $a^{-3/2}$, slower even than matter, because the decaying DM replenishes it.
Since in thermal equilibrium $T_d \propto \rho_\mathrm{DR}^{1/4}$, the attractor DR temperature drops as $T_d \propto a^{-3/8}$, much more slowly than decoupled radiation.

At our best-fit points, $\Gamma_d/H_0 \lesssim 10^{-4}$, the background expansion history is essentially indistinguishable from $\Lambda$CDM because the total fraction of DM that has decayed today is negligible ($10^{-4}$).
The DR contribution to the radiation energy density at recombination and earlier times is completely negligible 
and has no direct impact on big bang nucleosynthesis or the primary CMB damping tail.

\subsection{Perturbation equations}
\label{sec:perturbations}

We work in conformal Newtonian gauge with scalar metric potentials $\Phi$ and $\Psi$, and $\theta \equiv i k_j v^j$ is the velocity divergence.

Following the decaying-dark-matter perturbation framework of Refs.~\cite{Audren:2014bca,Poulin1}, the DM perturbations evolve according to
\begin{align}
\dot\delta_\chi &= -\theta_\chi + 3\dot\Phi - a\Gamma_d\,\Psi\,,\label{eq:dm_cont}\\
\dot\theta_\chi &= -\mathcal{H}\theta_\chi + k^2\Psi + a\Gamma_s\!\left(\theta_\mathrm{DR} - \theta_\chi\right).\label{eq:dm_euler}
\end{align}
Equation~\eqref{eq:dm_cont} is the standard continuity equation with an additional $-a\Gamma_d\,\Psi$ from the gravitational time dilation of decays in over-dense regions.
Equation~\eqref{eq:dm_euler} contains the drag $a\Gamma_s(\theta_\mathrm{DR} - \theta_\chi)$, which is new here; its physical origin and observational consequences are discussed in Sec.~\ref{sec:effects}.

Due to the strong self-interactions, the DR has no anisotropic stress and the Boltzmann hierarchy truncates. Its density and velocity perturbations obey
\begin{align}
\dot\delta_\mathrm{DR} &= -\frac{4}{3}\theta_\mathrm{DR} + 4\dot\Phi + a\Gamma_d\,\frac{\bar\rho_\chi}{\bar\rho_\mathrm{DR}}\!\left(\delta_\chi + \Psi - \delta_\mathrm{DR}\right),\label{eq:dr_cont}\\[6pt]
\dot\theta_\mathrm{DR} &= \frac{k^2}{4}\delta_\mathrm{DR} + k^2\Psi + \frac{3a\Gamma_d\bar\rho_\chi}{4\bar\rho_\mathrm{DR}}\!\left(\theta_\chi - \tfrac{4}{3}\theta_\mathrm{DR}\right)\nonumber\\
&\quad + \frac{3a\Gamma_s\bar\rho_\chi}{4\bar\rho_\mathrm{DR}}\!\left(\theta_\chi - \theta_\mathrm{DR}\right).\label{eq:dr_euler}
\end{align}
The $\Psi$ inside the parenthesis of Eq.~\eqref{eq:dr_cont} arises from the gravitational time-dilation of the proper decay rate in over-dense regions.
The energy density ratio $\bar\rho_\chi/\bar\rho_\mathrm{DR} \sim H/\Gamma_d \sim 10^4$ in Eqs.~\eqref{eq:dr_cont},\eqref{eq:dr_euler} strongly enhances the terms which couple DR to the DM and prevent the DR from oscillating.
In Eq.~\eqref{eq:dr_euler}, the decay energy injection term involves $\theta_\chi - \frac{4}{3}\theta_\mathrm{DR}$, where the $\frac{4}{3}$ reflects the $\rho+p$ normalization of the velocity perturbation for a relativistic fluid.
The scattering term involves $\theta_\chi - \theta_\mathrm{DR}$ without this factor, because scattering exchanges momentum rather than energy.
The prefactors $\frac{3\bar\rho_\chi}{4\bar\rho_\mathrm{DR}}$ ensure total momentum conservation, verifiable by combining Eqs.~\eqref{eq:dm_euler} and~\eqref{eq:dr_euler} with the appropriate density weights.
These perturbation equations are solved numerically using a modified version of \textsc{class}~\cite{CLASS}.

\section{Effects on Observables}
\label{sec:effects}

\subsection{Matter power spectrum}
\label{sec:pk_effect}

The DM--DR drag suppresses the growth of density perturbations on small scales while leaving large scales unaffected.
On sub-horizon scales, dark matter falls into gravitational potential wells, but the DR, supported by its own pressure gradient, resists compression and on sufficiently small scales streams out of DM-over-dense regions where it is produced from DM decay.
The elastic scattering transfers momentum between the in-falling DM and the out-flowing DR, producing the drag term $a\Gamma_s(\theta_\mathrm{DR} - \theta_\chi)$ in Eq.~\eqref{eq:dm_euler} which decelerates the DM and suppresses the growth of $\delta_\chi$.

The effectiveness of this drag depends on scale.
On scales larger than the drag horizon ($k \ll k_\mathrm{d}$), the coupling between the DR and DM dominates over the DR pressure, locking $\theta_\mathrm{DR} \approx \theta_\chi$.
The drag term therefore goes to zero and perturbations grow as in $\Lambda$CDM.
On small scales ($k \gg k_\mathrm{d}$), the DR pressure gradient is more important than the coupling, a relative velocity develops, and the drag suppresses $\delta_\chi$.
The crossover at $k_\mathrm{d}$ produces a step-shaped suppression in the matter power spectrum
\bea
\frac{P(k)}{P_{\Lambda\mathrm{CDM}}(k)} \simeq 1 - \frac{\mathcal{A}}{1+(k_d/k)^2}\,.
\eea

The effects of the two parameters $\Gamma_s$ and $\Gamma_d$ can be separated, in the relevant regime where $\Gamma_d \ll \Gamma_s \ll H_0$: $\Gamma_s$ controls the suppression depth, with $\mathcal{A} \propto \Gamma_s/H$ (Appendix~\ref{app:analytic}), while $\Gamma_d/\Gamma_s$ sets the step location through
\begin{equation}
	\label{eq:kstar}
	k_\mathrm{d} \sim \frac{15}{\tau}\sqrt{\frac{\Gamma_s}{\Gamma_d}} \,,
\end{equation}
where $\tau$ is the conformal time at the redshift of interest.
Reducing $\Gamma_d/\Gamma_s$ pushes $k_\mathrm{d}$ to larger $k$; in the limit $\Gamma_d/\Gamma_s \to 0$ the step moves beyond all observable scales and $\Lambda$CDM is recovered.
Setting $\Gamma_s = 0$ eliminates the drag entirely.

The time dependence of $\Gamma_s/H$ determines when the suppression develops.
For $n_{\Gamma_s} = 0$, $\Gamma_s/H$ grows as $a^{3/2}$ during matter domination; for $n_{\Gamma_s} = 2$, $\Gamma_s \propto T_d^2 \propto a^{-3/4}$ [Eq.~\eqref{eq:attractor}], giving $\Gamma_s/H \propto a^{3/4}$.
In both cases the suppression strengthens at late times.

\begin{figure}[t]
\centering
\includegraphics[width=\columnwidth]{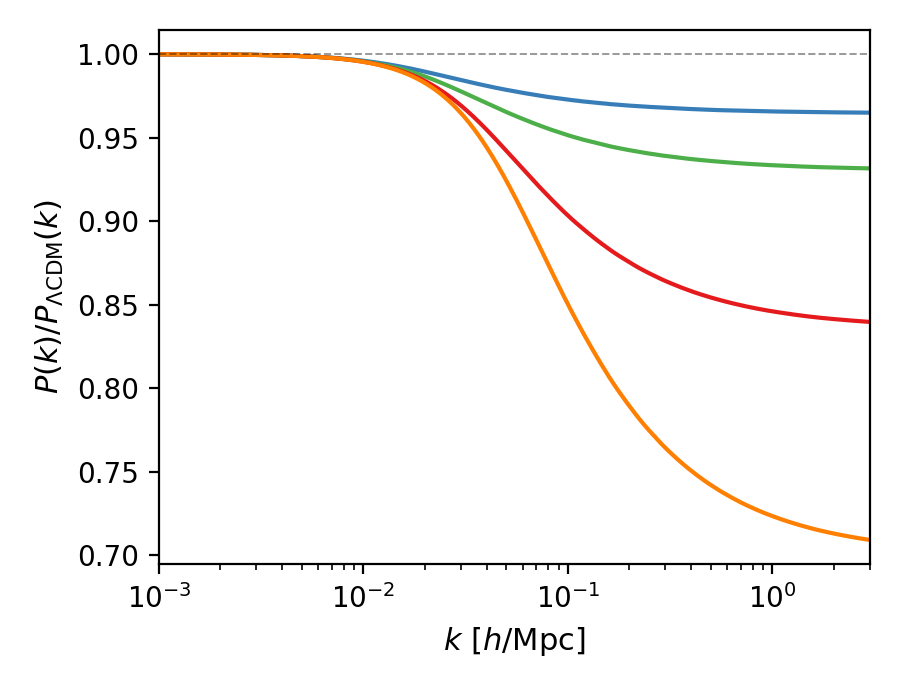}
\caption{$P(k)/P_{\Lambda\mathrm{CDM}}(k)$ at $z=0$ for $\Gamma_s^0/H_0 = 0.01$ (blue), $0.02$ (green), $0.05$ (red), $0.1$ (orange). All other parameters are held at the $n_{\Gamma_s}=2$ best-fit in Table~\ref{tab:bestfit_massless}. $P_{\Lambda\mathrm{CDM}}$ is computed with the same parameters, but with $\Gamma_s=0$.}
\label{fig:pk_Gs}
\end{figure}

\begin{figure}[t]
\centering
\includegraphics[width=\columnwidth]{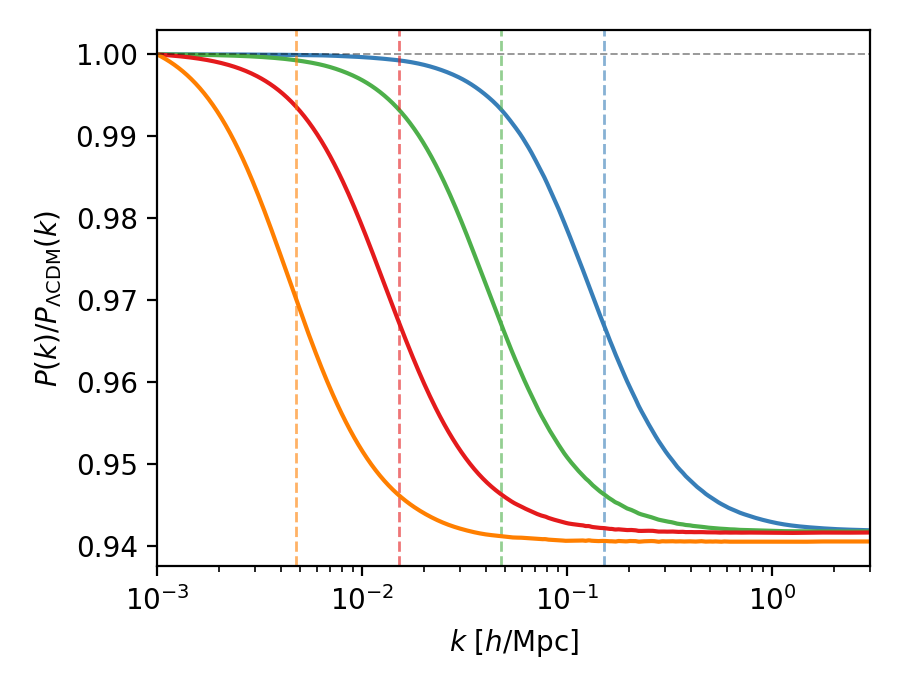}
\caption{$P(k)/P_{\Lambda\mathrm{CDM}}(k)$ at $z = 0$ with $n_{\Gamma_s} = 0$ for $\log_{10}(\Gamma_d/\Gamma_s) = -4$ (blue), $-3$ (green), $-2$ (red), $-1$ (orange). All other parameters are held at the $n_{\Gamma_s}=0$ best-fit in Table~\ref{tab:bestfit_massless}. $P_{\Lambda\mathrm{CDM}}$ is computed with the same parameters but with the drag turned off.
Dashed vertical lines mark $k_\mathrm{d}$ from Eq.~\eqref{eq:kstar}.}
\label{fig:pk_kstar}
\end{figure}

Figures~\ref{fig:pk_Gs} and~\ref{fig:pk_kstar} illustrate this separation: increasing $\Gamma_s$ deepens the step without shifting it, while decreasing $\Gamma_d/\Gamma_s$ shifts it to larger $k$ without changing its depth.

\subsection{Effect on the CMB}
\label{sec:cmb_effect}

At recombination, $\rho_\mathrm{DR}/\rho_\chi \sim \Gamma_d\,t_\mathrm{rec} \lesssim 10^{-6}$ (corresponding to $\Delta N_\mathrm{eff} \lesssim 10^{-5}$), and the DM--DR drag is still negligible, so the primary temperature and polarization spectra are unchanged.

Two late-time effects are modified, both small compared to current uncertainties in measurements. Since the drag suppresses the growth of $\delta_\chi$, it causes the gravitational potential $\Phi$ to evolve even during matter domination where it would otherwise be constant.
This modifies the late integrated Sachs-Wolfe signal in $C_\ell^{TT}$ at $\ell \lesssim 30$, though the effect is small because $\Gamma_s/H \ll 1$.
The suppression of $P(k)$ on small scales also reduces the CMB lensing potential $C_\ell^{\kappa\kappa}$, since lensing is an integrated measure of $P(k)$ weighted by a kernel peaking at $z \sim 1$--$2$.
Both effects are small at the best-fit values of $\Gamma_s$ and $\Gamma_d$ ($\sim 1\%$ for $n_{\Gamma_s} = 0$, $\sim 2.5\%$ for $n_{\Gamma_s} = 2$). The CMB therefore does not strongly constrain $\Gamma_s$ in this region of parameter space.

\subsection{Scale-dependent growth rate}
\label{sec:fkz}

In $\Lambda$CDM the linear growth rate $f(z) = d\ln\delta_m/d\ln a$ is independent of wavenumber for modes deep inside the horizon.
The step in $P(k)$ breaks this: modes with $k \gg k_\mathrm{d}$ experience the full drag and grow more slowly, while modes with $k \ll k_\mathrm{d}$ are unaffected.
At the best-fit parameters, the suppression of $f(k,z)/f_{\Lambda\mathrm{CDM}}(z)$ reaches approximately $5\%$ at $k \sim 1\;h$/Mpc and $z = 0$ for $n_{\Gamma_s} = 0$, and decreases at higher redshifts.

\begin{figure}[t]
\centering
\includegraphics[width=\columnwidth]{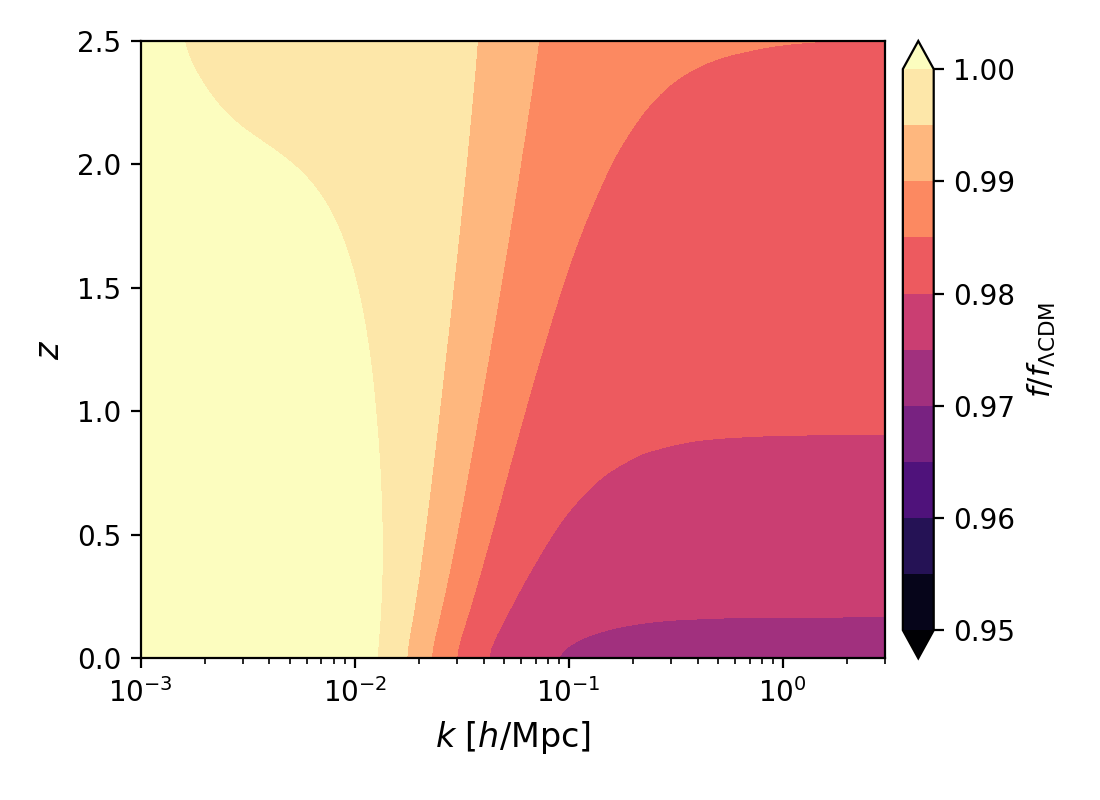}\\[6pt]
\includegraphics[width=\columnwidth]{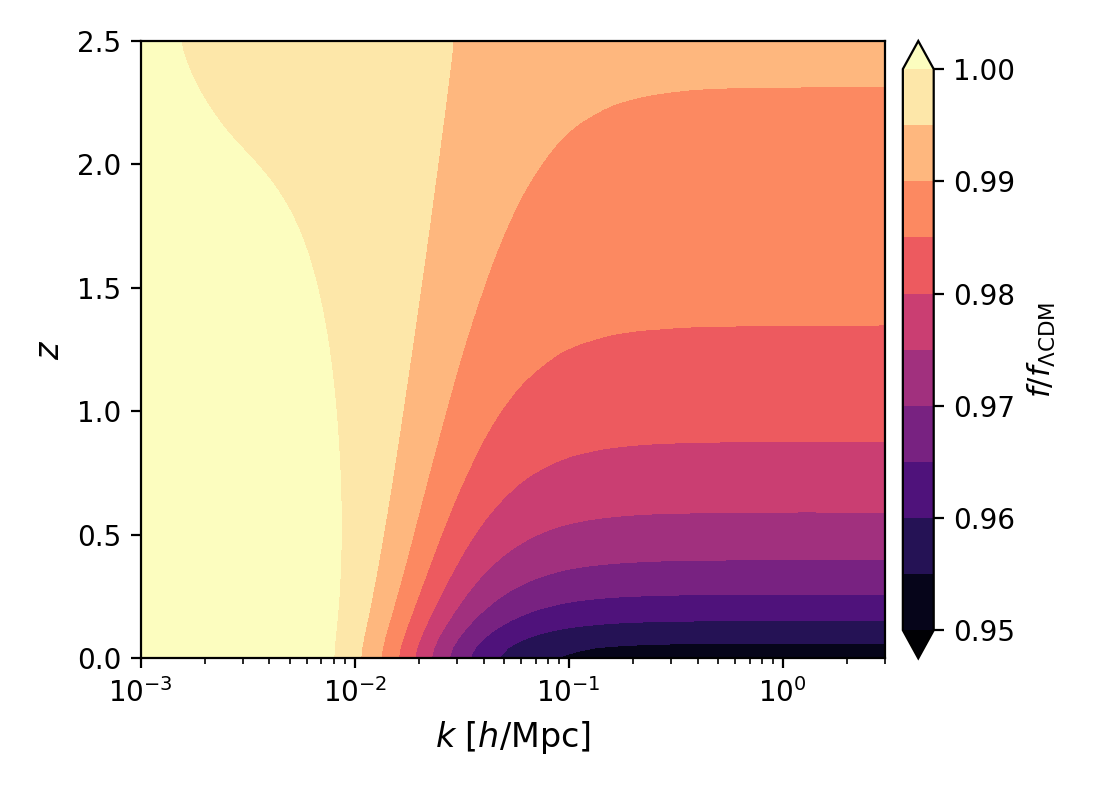}
\caption{$f(k,z)/f_{\Lambda\mathrm{CDM}}(z)$ at the best-fit points for $n_{\Gamma_s} = 2$ (top) and $n_{\Gamma_s} = 0$ (bottom).}
\label{fig:fkz}
\end{figure}

This scale-dependent growth rate, shown in Fig.~\ref{fig:fkz}, is a distinctive prediction of iDCDM.
$k$-resolved growth rate measurements at the $\sim 5\%$ level from DESI full-shape data, Euclid, or Rubin LSST could detect this signature.

The scale dependence of $f(k,z)$ has practical consequences for how $f\sigma_8$ is measured.
We construct the growth rate observable $f\sigma_8(z)$ from $P_\mathrm{cb}(k,z)$:
\begin{equation}
\label{eq:sigma8}
\sigma_8^2(z) = \int_0^\infty \frac{dk}{k}\,\frac{k^3}{2\pi^2}\,P_\mathrm{cb}(k,z)\,\left|W(kR_8)\right|^2,
\end{equation}
where $W(x) = 3(\sin x - x\cos x)/x^3$ is the Fourier-space top-hat window function at $R_8 = 8\;h^{-1}\mathrm{Mpc}$, and
\begin{equation}
\label{eq:fsigma8}
f\sigma_8(z) = -(1+z)\,\frac{d\sigma_8}{dz}\,,
\end{equation}
evaluated numerically within \textsc{class}.

The $f\sigma_8$ data are extracted from two types of surveys that probe the velocity field in distinct ways.
Redshift-space distortion (RSD) surveys measure the anisotropy in the galaxy two-point function induced by coherent peculiar velocities~\cite{KaiserModel}.
The resulting $f\sigma_8$ is obtained by fitting the redshift-space power spectrum to a $\Lambda$CDM-shaped template.
The relevant window function weighting the modes that contribute to the measurement is determined by the survey geometry and the $\sigma_8$ top-hat window $W(kR_8)$.

Peculiar velocity surveys, by contrast, measure galaxy velocities directly and extract $f\sigma_8$ from the velocity power spectrum.
The velocity field $\mathbf{v}(\mathbf{k}) \propto i\mathbf{k}\,\delta(\mathbf{k})/k^2$, so the velocity power spectrum carries an additional factor of $1/k^2$ relative to the density power spectrum.
The effective window function therefore differs from the density window, suppressing high-$k$ modes more strongly.
The two most constraining of the twenty $f\sigma_8$ measurements used in our analysis (Sec.~\ref{sec:datasets}) are derived from peculiar velocity surveys~\cite{Said_2020,Boruah_2020} that reconstruct the velocity field from the 2M++ density field~\cite{Carrick_2015} with Gaussian smoothing ($\sigma_s = 4\;h^{-1}\mathrm{Mpc}$).
The resulting velocity window peaks at $k \sim 0.07\;h/\mathrm{Mpc}$, well below the $\sigma_8$ top-hat window peak at $k \sim 0.2\;h/\mathrm{Mpc}$.

For $\Lambda$CDM, where the linear growth rate $f$ is independent of $k$, this distinction is immaterial: both windows yield the same $f\sigma_8$.
In iDCDM, however, the power spectrum suppression is $k$-dependent.
When $\Gamma_d/\Gamma_s$ is sufficiently small, the step in $P(k)/P_{\Lambda\mathrm{CDM}}(k)$ extends to scales where the density and velocity windows probe different effective growth rates.
In this regime the $f\sigma_8$ measurements from peculiar velocities become model-dependent and the standard $f\sigma_8$ likelihood, which assumes a $\Lambda$CDM-shaped template, is no longer applicable.

\begin{figure}[t]
\centering
\includegraphics[width=\columnwidth]{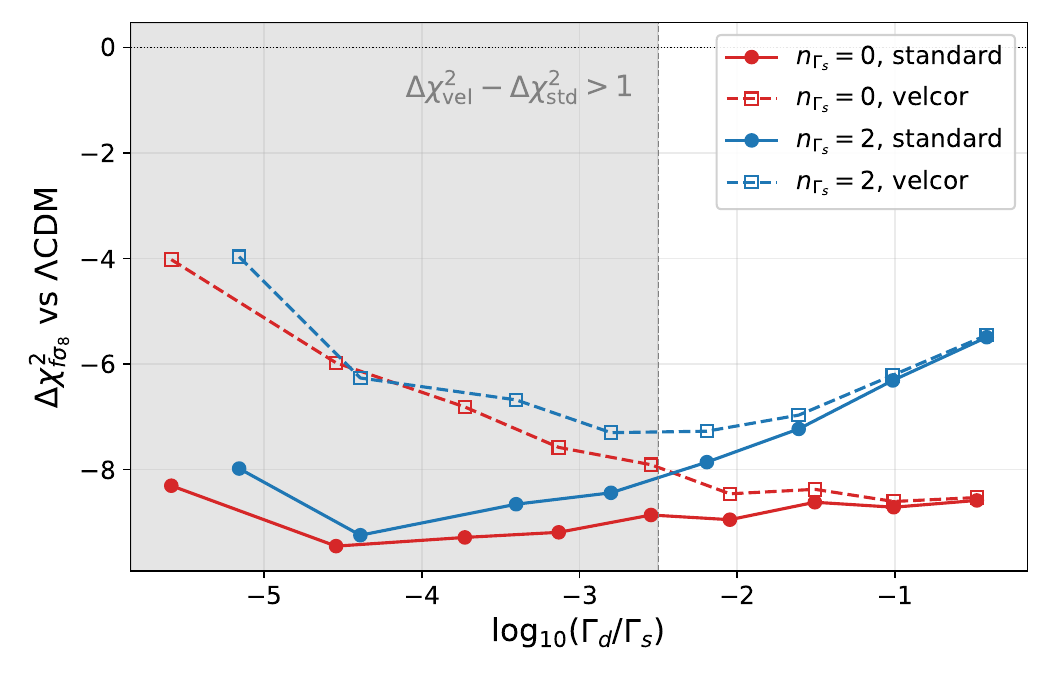}
\caption{Profile likelihood scan comparing the standard $f\sigma_8$ likelihood (``standard'') to the velocity-window-corrected likelihood (``velcor'') as a function of $\log_{10}(\Gamma_d/\Gamma_s)$. The shaded region marks $\log_{10}(\Gamma_d/\Gamma_s) < -2.5$, which we exclude as a prior boundary: for $\log_{10}(\Gamma_d/\Gamma_s) \gtrsim -2.5$ the two likelihoods agree to within $\Delta\chi^2 < 1$, while for smaller values the definition of $f\sigma_8$ becomes model-dependent and the two diverge. Non-monotonicity at the $\Delta\chi^2 \lesssim 1$ level reflects finite convergence of the Powell minimizer at each grid point.}
\label{fig:profile_fsig8}
\end{figure}

We determine where in our parameter space this model dependence becomes important 
by performing a profile likelihood scan over $\log_{10}(\Gamma_d/\Gamma_s)$, comparing the $\chi^2$ obtained with the standard $f\sigma_8$ likelihood to that obtained with a velocity-window-corrected likelihood that replaces the $\sigma_8$ window function with one that properly spans the scales to which the two most constraining peculiar velocity data points are sensitive.
Figure~\ref{fig:profile_fsig8} shows that for $\log_{10}(\Gamma_d/\Gamma_s) \gtrsim -2.5$, the two likelihoods agree to within $\Delta\chi^2 \lesssim 1$.
In order to remain insensitive to this model dependence, we impose $\log_{10}(\Gamma_d/\Gamma_s) > -2.5$ as a prior boundary.
We stress that it would be desirable to reanalyze the peculiar velocity surveys while allowing for $k$-dependent matter power spectrum templates and to test for such $k$-dependence in the data.

\section{Constraints from Current Data}
\label{sec:data}

\subsection{Datasets}
\label{sec:datasets}

We now explore what constraints current data already place on the iDCDM parameters. We use a combination of seven datasets spanning the CMB, large-scale structure, and distance measurements.

\emph{CMB Temperature and Polarization}: We use Planck 2018~\cite{Planck2018} high-$\ell$ TT+TE+EE, low-$\ell$ TT, and low-$\ell$ EE likelihoods with the full set of nuisance parameters.

\emph{CMB Lensing}: We include the combined ACT DR6~\cite{ACT,ACTDR6PS,APSLens3,APSLens4,APSLens5} + SPT-3G~\cite{APSLens2} + Planck PR4~\cite{APSLens6} lensing likelihood from Ref.~\cite{APSLens1}.

\emph{Baryon Acoustic Oscillations}: We use DESI DR2 BAO measurements~\cite{DESIDR2}.

\emph{Type Ia Supernovae}: We use the uncalibrated Pantheon+ compilation~\cite{Pantheon+}.

\emph{Growth Rate}: We use twenty $f\sigma_8$ measurements compiled from redshift-space distortion and peculiar velocity surveys spanning $z = 0$ to $z \simeq 1.4$, as assembled in Ref.~\cite{Huterer1} (see also Refs.~\cite{DESIfsig8,Huterer2}).
The individual measurements are drawn from 6dFGS~\cite{Beutler_2012}, SDSS~\cite{Howlett_2015}, GAMA~\cite{Blake_2013}, WiggleZ~\cite{Blake_2012}, VIPERS~\cite{Pezzotta_2017}, FastSound~\cite{Okumura_2016}, and peculiar velocity surveys~\cite{Huterer_2017,Said_2020,Boruah_2020,Turner_2022,Carrick_2015}.
We treat the measurement errors as uncorrelated, following the convention of the referenced compilations.
The twenty measurements are listed in Table~\ref{tab:fsig8_data}.

\begin{table}[t]
  \caption{$f\sigma_8$ data used in the growth rate likelihood. Measurements marked with an asterisk ($^\ast$) are derived from peculiar-velocity surveys; the remaining points are RSD measurements. Classification follows~\cite{Huterer1}.}
  \label{tab:fsig8_data}
  \begin{ruledtabular}
  \begin{tabular}{cccc}
  $z$ & $f\sigma_8$ & $z$ & $f\sigma_8$ \\
  \hline
  $0.02$  & $0.400 \pm 0.017^\ast$ & 0.44 & $0.413 \pm 0.080$ \\
  $0.02$  & $0.428 \pm 0.046^\ast$ & 0.51 & $0.455 \pm 0.039$ \\
  $0.025$ & $0.358 \pm 0.075^\ast$ & 0.60 & $0.390 \pm 0.063$ \\
  $0.035$ & $0.338 \pm 0.027^\ast$ & 0.60 & $0.550 \pm 0.120$ \\
  0.067 & $0.423 \pm 0.055$ & 0.70 & $0.448 \pm 0.043$ \\
  0.15  & $0.490 \pm 0.150$ & 0.73 & $0.437 \pm 0.072$ \\
  0.15  & $0.530 \pm 0.160$ & 0.85 & $0.315 \pm 0.095$ \\
  0.18  & $0.360 \pm 0.090$ & 0.86 & $0.400 \pm 0.110$ \\
  0.38  & $0.440 \pm 0.060$ & 1.40 & $0.494 \pm 0.123$ \\
  0.38  & $0.500 \pm 0.047$ & 1.48 & $0.462 \pm 0.045$ \\
  \end{tabular}
  \end{ruledtabular}
  \end{table}

\subsection{Analysis pipeline and parameters}
\label{sec:priors}

We perform MCMC sampling using \textsc{MontePython}~\cite{Montepython1,Montepython2} interfaced with a modified version of \textsc{class} v3.3.4~\cite{CLASS}.
The modifications implement the iDCDM background [Eqs.~\eqref{eq:bg_chi}--\eqref{eq:bg_DR}] and perturbation [Eqs.~\eqref{eq:dm_cont}--\eqref{eq:dr_euler}] equations.

The $\Lambda$CDM sector is described by six parameters with flat priors:
$\{\omega_b,\, \omega_\mathrm{dm},\, h, \ln(10^{10}A_s),\, n_s,\, \tau_\mathrm{reio}\}$.
For \mbox{iDCDM}, $\omega_\mathrm{dm}$ is the dark matter density today that one would obtain in the absence of decay (i.e., for the same initial conditions, evolved with pure $a^{-3}$ dilution). The actual DM density today is less than $\omega_\mathrm{dm}$ because of the decay.

The iDCDM extension adds two free parameters, both evaluated at $z=0$:
\begin{itemize}
\item $\Gamma_s^0\equiv\Gamma_s(z=0)$: the DM--DR scattering rate today, with a linear non-negative prior ($\Gamma_s^0\geq 0$). We quote the natural dimensionless parameter, $\Gamma_s^0/H_0$.
\item $\log_{10}(\Gamma_d/\Gamma_s^0)$: the decay-to-scattering ratio at $z=0$, with a flat prior and a lower boundary of $-2.5$. This boundary is required in order to remain within the region of validity of the $f\sigma_8$ data, as discussed in Sec.~\ref{sec:fkz}.
\end{itemize}
Throughout the rest of the paper we use $\Gamma_s/H_0$ for $\Gamma_s^0/H_0$ and $\Gamma_d/\Gamma_s$ for $\Gamma_d/\Gamma_s^0$ when no confusion with the time-dependent rate arises.
We present results for two choices of the scattering rate scaling: $n_{\Gamma_s} = 0$ (constant rate) and $n_{\Gamma_s} = 2$ (rate obtained in the explicit particle physics model, $\Gamma_s \propto T_d^2$).
Our default analysis assumes massless neutrinos; the impact of a nonzero neutrino mass is discussed at the end of this section.

Non-linear corrections use the \textsc{halofit} prescription~\cite{Smith:2002dz,Takahashi:2012em}. As a cross-check, we recompute the non-linear corrections with \textsc{hmcode}~\cite{Mead:2015yca,Mead:2020vgs} and obtain a lensing $\chi^2$ shifted by only one unit, approximately constant across $\Gamma_s$, so the choice of non-linear prescription does not affect our parameter-space constraints.
Note that the $f\sigma_8$ likelihood is independent of the non-linear prescription since it uses $\sigma_8$ from the linear power spectrum.

\subsection{The Fit}
\label{sec:results}

Tables~\ref{tab:bestfit_massless} and~\ref{tab:chi2_massless} summarize the best-fit parameters and per-likelihood $\chi^2$ for massless neutrinos.
The \mbox{iDCDM} model has two parameters beyond $\Lambda$CDM: $\Gamma_s\ge 0$ and $\log_{10}(\Gamma_d/\Gamma_s)\ge -2.5$.

\begin{figure}[t]
\centering
\includegraphics[width=\columnwidth]{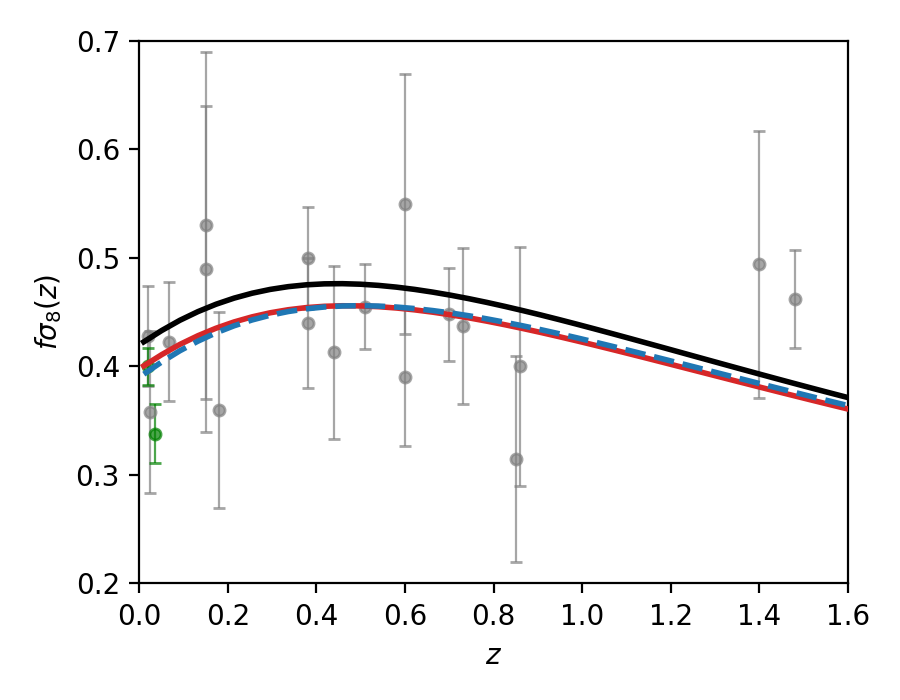}
\caption{$f\sigma_8(z)$ at the best-fit parameters compared to data (gray points with error bars).
$\Lambda$CDM best fit (black solid), iDCDM $n_{\Gamma_s} = 2$ best fit (red solid), iDCDM $n_{\Gamma_s} = 0$ best fit (blue dashed).
Both iDCDM models produce a lower $f\sigma_8$ at low redshift, driven by the step-shaped suppression of $P_\mathrm{cb}(k)$. The two green data points correspond to the peculiar velocity survey data which drive most of the tension with $\Lambda$CDM in the fit.}
\label{fig:fsig8_data}
\end{figure}

We find an improvement in $\chi^2$ relative to $\Lambda$CDM corresponding to about $2\sigma$ for both scalings, slightly stronger for $n_{\Gamma_s} = 0$: $\Delta\chi^2 = -7.6$ for $n_{\Gamma_s} = 0$ and $\Delta\chi^2 = -5.6$ for $n_{\Gamma_s} = 2$.
The improvement is driven almost entirely by the growth rate data: $\chi^2_{f\sigma_8}$ drops from 25.5 in $\Lambda$CDM to 16.9 in the $n_{\Gamma_s} = 0$ model ($\Delta\chi^2_{f\sigma_8} = -8.6$), slightly larger in magnitude than the total $\Delta\chi^2 = -7.6$ because the remaining likelihoods worsen by a net $\sim 1$, dominated by BAO. Fig.~\ref{fig:fsig8_data} shows the best-fit model's predictions for $f\sigma_8$ as a function of redshift compared with the data points making up $\chi^2_{f\sigma_8}$. It is clear that the improvement for iDCDM comes from the better fit to the peculiar velocity survey data points highlighted in green.
The BAO $\chi^2$ increases modestly ($+1.8$ for $n_{\Gamma_s} = 0$), driven by the upward shift in $\Omega_m$ from 0.297 to 0.302 that slightly modifies the distance-redshift relation.

\begin{table}[t]
\caption{Best-fit parameters for $\Lambda$CDM and the two iDCDM models from a fit to the seven-likelihood combination and assuming massless neutrinos. Note that the best-fit value for $\log_{10}(\Gamma_d/\Gamma_s)$ lies at the prior boundary $-2.5$ in both iDCDM models.}
\label{tab:bestfit_massless}
\begin{ruledtabular}
\begin{tabular}{lccc}
Parameter & $\Lambda$CDM & $n_{\Gamma_s}=2$ & $n_{\Gamma_s}=0$ \\
\hline
$100\,\omega_b$ & 2.257 & 2.248 & 2.247 \\
$\omega_\mathrm{dm}$ & 0.1181 & 0.1189 & 0.1188 \\
$100\,\theta_s$ & 1.04200 & 1.04200 & 1.04200 \\
$\ln(10^{10}A_s)$ & 3.054 & 3.054 & 3.055 \\
$n_s$ & 0.971 & 0.970 & 0.970 \\
$\tau_\mathrm{reio}$ & 0.060 & 0.060 & 0.060 \\
$\Gamma_s^0/H_0$ & --- & 0.023 & 0.053 \\
\hline
$H_0$ & 68.8 & 68.5 & 68.5 \\
$\Omega_m$ & 0.297 & 0.302 & 0.302 \\
$\sigma_8$ & 0.82 & 0.80 & 0.80 \\
\end{tabular}
\end{ruledtabular}
\end{table}

\begin{table}[t]
\caption{$\chi^2$ comparison with massless neutrinos.}
\label{tab:chi2_massless}
\begin{ruledtabular}
\begin{tabular}{lccc}
Likelihood & $\Lambda$CDM & $n_{\Gamma_s}=2$ & $n_{\Gamma_s}=0$ \\
\hline
Planck high-$\ell$ TTTEEE & 2346.1 & 2346.0 & 2345.8 \\
Planck low-$\ell$ TT & 22.5 & 22.8 & 22.9 \\
Planck low-$\ell$ EE & 397.2 & 397.2 & 397.4 \\
ACT+SPT+Planck lensing & 37.9 & 38.8 & 37.9 \\
Pantheon+ & 1414.0 & 1412.9 & 1413.0 \\
DESI DR2 BAO & 10.5 & 12.4 & 12.3 \\
$f\sigma_8$ & 25.5 & 18.0 & 16.9 \\
\hline
Total & 4253.7 & 4248.1 & 4246.1 \\
$\Delta\chi^2$ & --- & $-5.6$ & $-7.6$ \\
\end{tabular}
\end{ruledtabular}
\end{table}

The combination $A_s\,e^{-2\tau}$ is very well constrained by Planck. Separately, the fit shifts $\ln(10^{10}A_s)$ by $\lesssim 0.001$ between $\Lambda$CDM and iDCDM. Thus the $\sigma_8$ reduction is produced by the drag, not by a change in the primordial amplitude.

\begin{figure}[t]
\centering
\includegraphics[width=\columnwidth]{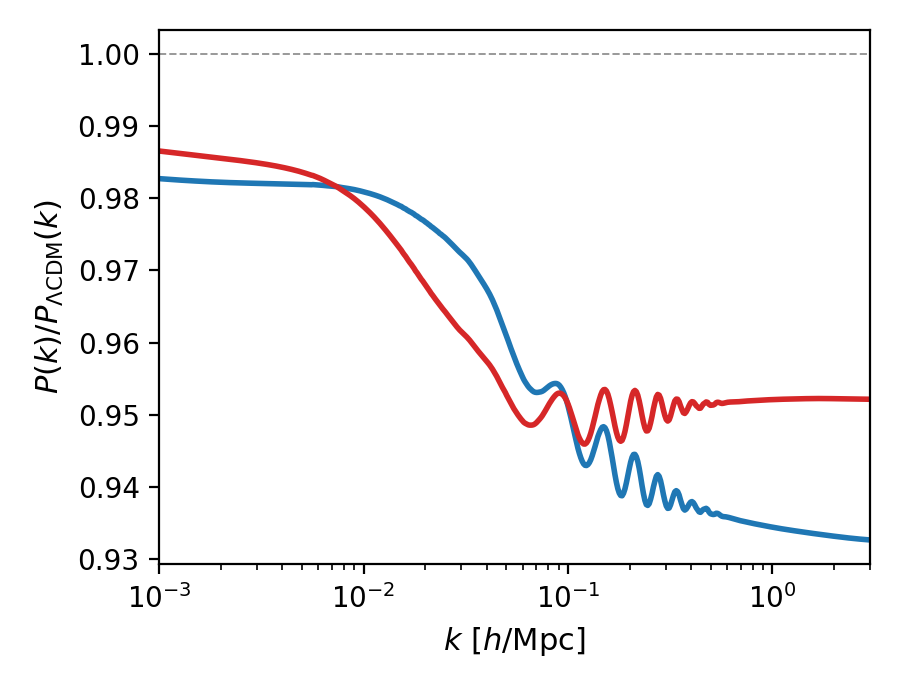}
\caption{Ratio of the best-fit iDCDM linear matter power spectrum to the best-fit $\Lambda$CDM at $z = 0$, for $n_{\Gamma_s} = 0$ (red) and $n_{\Gamma_s} = 2$ (blue).
For both model variants, the step-shaped suppression leaves scales $k \ll k_\mathrm{d}$ largely unaffected and reduces power at $k \gg k_\mathrm{d}$. The BAO wiggles in both cases are due to small changes in the background parameters.}
\label{fig:pk_shape}
\end{figure}

Both iDCDM best fits have $\log_{10}(\Gamma_d/\Gamma_s)$ at the prior boundary of $-2.5$ (Sec.~\ref{sec:fkz}), placing the $z=0$ step at $k_\mathrm{d} \sim 0.03\;h$/Mpc [Eq.~(\ref{eq:kstar}) and see Fig.~\ref{fig:pk_shape}], and reducing $\sigma_8$ from $0.82$ to $\sim 0.80$. The fact that the best-fit lies at the prior boundary shows that the data favor pushing $k_\mathrm{d}$ to large wavenumber (i.e., $\Gamma_d/\Gamma_s\to$ small) to decrease the range of suppressed modes.
We do not attempt a careful statistical interpretation of the (modest) current $\Delta\chi^2$ differences because of the absence of covariances between overlapping data points and the model dependence of the $f\sigma_8$ likelihood. We briefly discuss the implications of our best-fit points sitting at the $\Gamma_d/\Gamma_s$ boundary in Appendix~\ref{app:stats}.

The results above assume massless neutrinos.
Since massive neutrinos also suppress small-scale power through free-streaming, the treatment of neutrino masses affects how much room remains for the DM--DR drag to improve the $f\sigma_8$ fit. To investigate this, we
repeat the analysis with a single massive neutrino species with a mass of $0.06$~eV.
The best-fit parameters and $\chi^2$ breakdown are given in Tables~\ref{tab:bestfit_massive} and~\ref{tab:chi2_massive}.

\begin{table}[t]
\caption{Best-fit parameters assuming one massive neutrino with $m_\nu = 0.06$~eV. Same datasets and priors as Table~\ref{tab:bestfit_massless}. Again, the best-fit satisfies $\log_{10}(\Gamma_d/\Gamma_s)=-2.5$ in both iDCDM models.}
\label{tab:bestfit_massive}
\begin{ruledtabular}
\begin{tabular}{lccc}
Parameter & $\Lambda$CDM & $n_{\Gamma_s}=2$ & $n_{\Gamma_s}=0$ \\
\hline
$100\,\omega_b$ & 2.255 & 2.254 & 2.252 \\
$\omega_\mathrm{dm}$ & 0.1179 & 0.1184 & 0.1183 \\
$100\,\theta_s$ & 1.04210 & 1.04210 & 1.04210 \\
$\ln(10^{10}A_s)$ & 3.056 & 3.057 & 3.056 \\
$n_s$ & 0.972 & 0.971 & 0.971 \\
$\tau_\mathrm{reio}$ & 0.061 & 0.061 & 0.061 \\
$\Gamma_s^0/H_0$ & --- & 0.017 & 0.040 \\
\hline
$H_0$ & 68.4 & 68.2 & 68.2 \\
$\Omega_m$ & 0.302 & 0.304 & 0.304 \\
$\sigma_8$ & 0.81 & 0.79 & 0.79 \\
\end{tabular}
\end{ruledtabular}
\end{table}

\begin{table}[t]
\caption{$\chi^2$ comparison with one massive neutrino, $m_\nu = 0.06$~eV.}
\label{tab:chi2_massive}
\begin{ruledtabular}
\begin{tabular}{lccc}
Likelihood & $\Lambda$CDM & $n_{\Gamma_s}=2$ & $n_{\Gamma_s}=0$ \\
\hline
Planck high-$\ell$ TTTEEE & 2348.1 & 2347.8 & 2347.9 \\
Planck low-$\ell$ TT & 22.3 & 22.7 & 22.6 \\
Planck low-$\ell$ EE & 397.6 & 397.7 & 397.7 \\
ACT+SPT+Planck lensing & 40.0 & 40.9 & 40.5 \\
Pantheon+ & 1412.9 & 1412.4 & 1412.4 \\
DESI DR2 BAO & 12.2 & 14.4 & 14.3 \\
$f\sigma_8$ & 24.2 & 18.8 & 17.5 \\
\hline
Total & 4257.4 & 4254.7 & 4252.9 \\
$\Delta\chi^2$ & --- & $-2.7$ & $-4.5$ \\
\end{tabular}
\end{ruledtabular}
\end{table}

With massive neutrinos, the preference for iDCDM weakens: $\Delta\chi^2 = -4.5$ for $n_{\Gamma_s} = 0$ and $-2.7$ for $n_{\Gamma_s} = 2$. The reduced preference for iDCDM comes from two sets of data.

First, neutrino free-streaming partially suppresses $\sigma_8$.
In $\Lambda$CDM the best-fit $\sigma_{8,\mathrm{cb}}$~\footnote{Galaxy clustering observables are sensitive to perturbations in the clustering matter $\delta_\mathrm{cb}=(\rho_{\rm cdm} \delta_{\rm cdm}+\rho_b \delta_b)/(\rho_{\rm cdm}+\rho_b)$ not including massive neutrinos (or any other non-clustering matter)~\cite{Castorina:2013wga,Costanzi:2013bha}.} drops from 0.82 (massless) to 0.81 (massive).
With $\sigma_{8,\mathrm{cb}}$ already closer to the $f\sigma_8$ data, there is less room for the DM--DR drag to improve the fit.
The $f\sigma_8$ improvement weakens accordingly: $\Delta\chi^2_{f\sigma_8}$ shrinks from $-8.6$ to $-6.7$ for $n_{\Gamma_s} = 0$.

Massive neutrinos and the DM--DR drag in iDCDM both reduce the CMB lensing potential. Since current CMB lensing data prefer slightly more lensing power than the $\Lambda$CDM prediction, the lensing $\chi^2$ worsens by about 1 for iDCDM with massive neutrinos. 

Together, the statistical preference for iDCDM is reduced: the improvement of the $\chi^2$ for $f\sigma_8$ is smaller, and the CMB lensing fit is marginally degraded.

The full posterior distributions are shown in Appendix~\ref{app:posteriors}.
As expected from the model's construction, the triangle plot (Fig.~\ref{fig:full_triangle}) shows minimal correlations between the iDCDM and standard $\Lambda$CDM parameters.
Since $\Gamma_d/H_0 \lesssim 10^{-4}$ and the scattering affects only the perturbation equations, the standard $\Lambda$CDM parameters ($\omega_b$, $100\,\theta_s$, $n_s$, $\tau_\mathrm{reio}$) are determined by the CMB primary anisotropies and show very little correlation with $\Gamma_s$ or $\log_{10}(\Gamma_d/\Gamma_s)$.
This confirms that the iDCDM sector approximately decouples from the standard cosmological parameters.

\begin{figure}[t]
	\centering
	\includegraphics[width=\columnwidth]{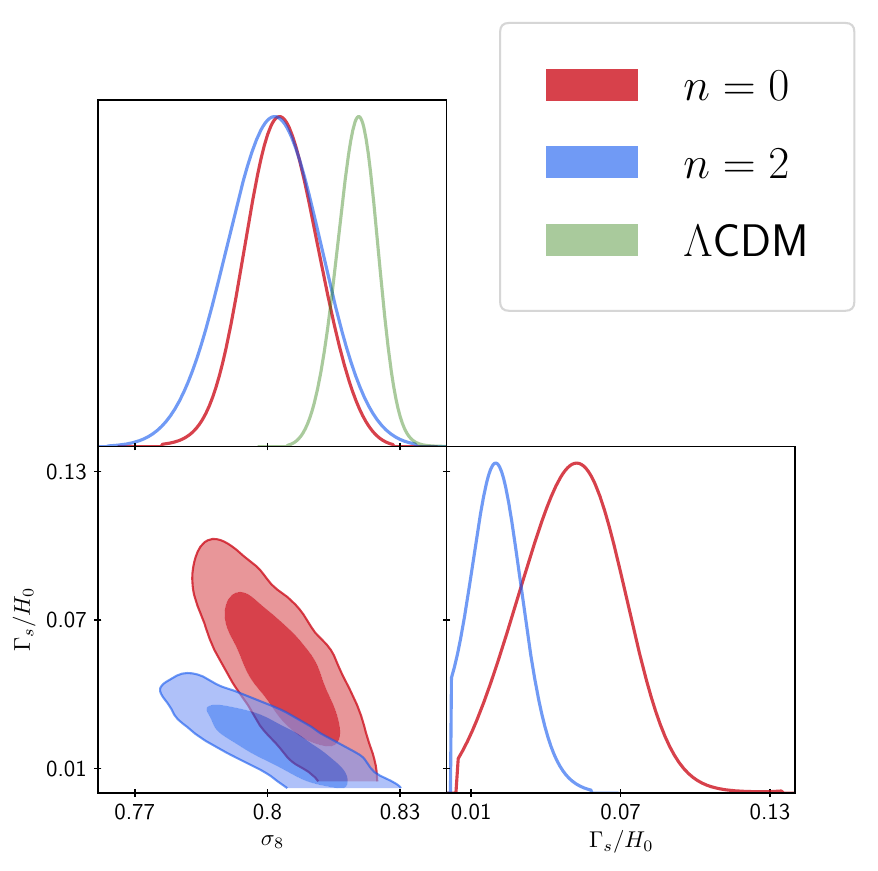}
	\caption{Two-dimensional marginalized posterior distributions for $\sigma_8$ and $\Gamma_s^0/H_0$, with massless neutrinos.}
	\label{fig:smalltriangle}
\end{figure}

The few panels that do exhibit visible correlations carry physical information.
The iDCDM parameters $\Gamma_s$ and $\log_{10}(\Gamma_d/\Gamma_s)$ are correlated with each other because a deeper step (larger $\Gamma_s$) can be partially compensated at the scales relevant for $f\sigma_8$ by pushing the step to larger $k$ (decreasing $\Gamma_d/\Gamma_s$).
The parameter $\sigma_8$ is anti-correlated with $\Gamma_s$, since increasing the scattering rate suppresses the amplitude of the matter power spectrum (Fig.~\ref{fig:smalltriangle}). We also see that the posteriors of the $n_{\Gamma_s} = 0$ model are shifted to larger values of $\Gamma_s$ than those for $n_{\Gamma_s} = 2$. This is because $\Gamma_s$ is the momentum transfer rate at $z=0$ and for $n_{\Gamma_s} = 0$ the momentum transfer rate rises more steeply at late times, thus the suppression of growth acts only over a short range of redshifts and $\Gamma_s$ must be larger to compensate.  
\begin{figure}[t]
\centering
\includegraphics[width=\columnwidth]{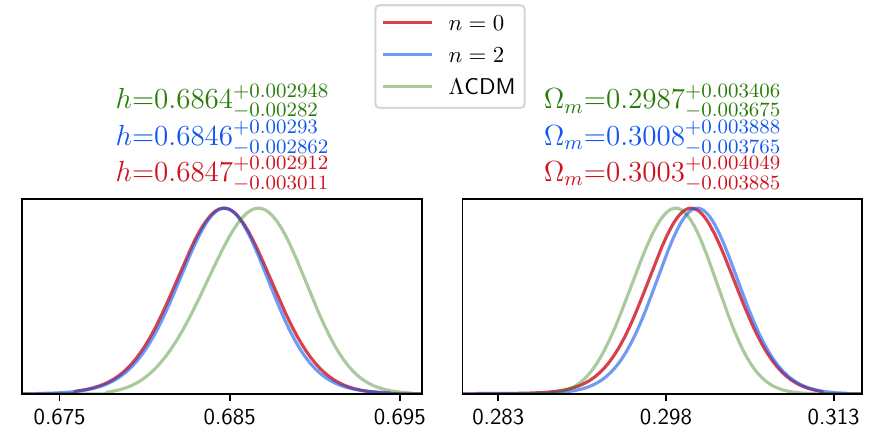}
\caption{One-dimensional marginalized posterior distributions for $h$ and $\Omega_m$, with massless neutrinos.}
\label{fig:small1d_posteriors}
\end{figure}

The one-dimensional posteriors (Figs.~\ref{fig:small1d_posteriors} and~\ref{fig:1d_posteriors_full}) confirm that the standard $\Lambda$CDM parameters are consistent across the three fits.
The only parameter whose posterior shifts appreciably when iDCDM is turned on is $\sigma_8$. The $\Lambda$CDM parameters $\omega_\mathrm{dm}$, $h$, $\tau$ (due to a modified late integrated Sachs-Wolfe), and $A_s$ (correlated with $\tau$) shift by less than 1$\sigma$.

\section{Conclusions}
\label{sec:conclusions}

Motivated by recent LSS data which suggest that the growth of structure may be suppressed at low redshift, we have discussed a class of models in which dark matter experiences a drag against dark radiation that grows at late times. As an explicit example we presented iDCDM, a model with an unbroken dark $U(1)$ gauge group. Dark matter and a massless fermion are both charged under the $U(1)$ and scatter elastically through dark photon exchange. Dark matter decays into the massless fermion and the dark photon which together form an interacting DR bath. The decay continuously replenishes the DR bath, allowing the ratio $\Gamma_s/H$ that controls the drag to grow with time rather than decaying, the opposite of conventional DM--DR models.

The two new parameters of iDCDM map directly onto observable features: $\Gamma_s$ sets the suppression depth $\mathcal{A}$, while $\Gamma_d/\Gamma_s$ sets the step location $k_\mathrm{d}$ [Eq.~\eqref{eq:kstar}].
The background expansion, BBN, and primary CMB are untouched ($\Gamma_d/H_0\lesssim 10^{-4}$); the model acts primarily on perturbations.
  
The testable prediction is a step-shaped, scale-dependent linear growth rate $f(k,z)$ with transition at $k_\mathrm{d}$: modes with $k\gg k_\mathrm{d}$ grow more slowly, while modes with $k\ll k_\mathrm{d}$ are unaffected.
No scale-independent modification of growth (rescalings of $\sigma_8$, $\Lambda$CDM retuning, or smooth dark-energy variations) can produce this profile; the $k$-dependence combined with the predicted $z$-evolution is the model's fingerprint.
Massive neutrino free-streaming also suppresses small-scale power but the suppression starts at much higher redshifts when neutrinos become non-relativistic and has a different $k$-profile, so the two effects are physically separable given sufficient $k$- and $z$-resolution. For parameters near the current best fit, the suppression of $f(k,z)/f_{\Lambda\mathrm{CDM}}(z)$ reaches $\sim 5\%$ at $k\sim 1\,h$/Mpc, $z=0$, a concrete target for upcoming surveys.

In a first confrontation with current data (Planck 2018, ACT+SPT+Planck lensing, DESI DR2 BAO, Pantheon+, and 20 $f\sigma_8$ measurements), both scattering-rate scalings show a modest preference for the drag signature, with $\Delta\chi^2 = -7.6$ ($n_{\Gamma_s}=0$) and $-5.6$ ($n_{\Gamma_s}=2$) at best-fit values $\Gamma_s^0/H_0 = 0.053$ and $0.023$, respectively. The improvement is driven by $f\sigma_8$; the CMB, lensing, BAO, and supernova fits are essentially unchanged. The resulting $\sigma_8$ reduction is produced by the drag itself rather than by a shift in the primordial amplitude ($\Delta\ln(10^{10}A_s)\lesssim 10^{-3}$). With one massive neutrino ($m_\nu = 0.06$~eV), drag and free-streaming both suppress small-scale power and reduce the CMB lensing prediction relative to the data; the preferences shrink to $\Delta\chi^2 = -4.5$ and $-2.7$. Current data are not decisive; we speculate that whatever physics resolves the present preference of global fits for negative neutrino masses~\cite{Craig:2024tky,Green:2024xbb,Graham:2025dqn} may also have a significant impact on the goodness of fit for iDCDM.

Similar realizations of this physics can be constructed. For example,~\cite{Dallari:2026wvj} suggests that a theory with scalars and trilinear couplings may yield a late-time-growing drag from scattering off primordial DR (redshifting as $T_{\rm DR} \propto a^{-1}$) at the expense of fine-tuning; iDCDM achieves it instead through decay-replenished DR with technically natural couplings and masses. Their analysis of CMB + DESI BAO data places only an upper bound, $\Gamma_s^0/H_0\lesssim 0.04$. Their physical DM drag rate scales with redshift as $a^{-1}$ which is close to our $n_{\Gamma_s}=2$ model where the physical DM drag rate scales as $a^{-3/4}$. Therefore the relevant comparison is to our $n_{\Gamma_s}=2$ best fit, $\Gamma_s^0/H_0 = 0.023$, which is consistent with their bound; the modest preference we find is driven by the $f\sigma_8$ data their analysis does not include.

 Alternatively, one may ask if iDCDM could be modeled by choosing appropriate ETHOS parameters because~\cite{Dallari:2026wvj} also find a step-like suppression of the matter power spectrum. There are two differences: {\it i.} DR produced from decay in iDCDM streams {\it out of} overdense regions with a velocity $15/8$ times the infall velocity of DM so that the relative velocity entering the drag term is $23/8$ times larger in iDCDM. To partially compensate, an ETHOS model would have its scattering rate boosted by $23/8$. {\it ii.} More significantly, the background radiation density redshifts much more quickly in ETHOS than in iDCDM. Therefore, in order to obtain the same radiation density and associated location of the step in the matter power spectrum at $z=0$, one must start with $\sim 10^7$ times larger radiation densities at recombination (larger $N_{\rm eff}$) in ETHOS, which can affect the primary CMB perturbations. For example, approximately matching the amplitude and $k$-dependence of our iDCDM best fit with $n_{\Gamma_s}=0$ requires choosing $\Delta N_{\rm eff}\simeq0.2$ in ETHOS, while the iDCDM best fit with $n_{\Gamma_s}=2$ requires $\Delta N_{\rm eff}\simeq0.02$.

DESI full-shape spectroscopy, Euclid, and Rubin LSST weak lensing can reach the $\sim 5\%$ sensitivity on $f(k,z)$ needed to detect or exclude the predicted step at the current best-fit amplitude. We suggest that future large scale structure data be fit not only with $\Lambda$CDM but also with models which predict departures of $P(k,z)$ from simple power laws in both $z$ and $k$ such as iDCDM. Such fits would also extend the reach of our analysis to smaller $\Gamma_d/\Gamma_s$, which is currently limited by the $\Lambda$CDM template assumed in the published $f\sigma_8$ measurements.

\begin{acknowledgments}
This research is supported by DOE Award DE-SC0015845.
M.S. thanks Prof.\ Laura Covi at G\"ottingen U.\ for her hospitality and support and Dr.\ Sarif Khan for collaboration during the initial stages of this project.
Our MCMC runs were performed on the Shared Computing Cluster, which is administered by Boston University's Research Computing Services. Claude (Anthropic) was used as a research and writing assistant during the preparation of this manuscript; the authors are responsible for all scientific content.
\end{acknowledgments}

\appendix

\section{Analytical Approximation}
\label{app:analytic}

The numerical results presented in the main text are obtained by solving the full coupled perturbation equations within \textsc{class}.
In order to gain more intuition, we also derive an analytical formula for the power spectrum suppression $P_m(k)/P_{\Lambda\mathrm{CDM}}(k)$ that reproduces the numerical output to an accuracy of a few percent.
The derivation proceeds perturbatively in the small parameters $\Gamma_s/H$ and $\Gamma_d/H$, assumes matter domination, and is carried out for $n_{\Gamma_s} = 0$ (constant scattering rate).

The restriction to $n_{\Gamma_s} = 0$ is not merely for mathematical convenience.
The key physical requirement is that the scattering-induced suppression be dominated by late-time contributions, so that an expansion around the standard $\Lambda$CDM growing mode captures the leading effect.
For $n_{\Gamma_s} = 0$, $\Gamma_s$ is constant and $\Gamma_s/H \propto a^{3/2}$ during matter domination, growing strongly with time.
The drag therefore becomes progressively more important, and the late-time contribution dominates the integrated effect, permitting a clean perturbative solution.
For $n_{\Gamma_s} = 2$, $\Gamma_s \propto T_d^2 \propto a^{-3/4}$ [Eq.~\eqref{eq:attractor}], so $\Gamma_s/H \propto a^{3/4}$, still growing but not strongly enough to ensure that all integrals of the perturbation equations are dominated at late times. 
The sourced growth equation in this case does not admit a simple closed-form solution.

\subsection{Background during matter domination}

During matter domination the DR density is governed by the attractor solution $\bar\rho_\mathrm{DR} \approx \frac{2}{5}\bar\rho_\chi\,\Gamma_d/H$ [Eq.~\eqref{eq:attractor}].
Thus we may approximate the ratio
\begin{equation}
\label{eq:R_def}
 \frac{3\bar\rho_\chi}{4\bar\rho_\mathrm{DR}} \approx \frac{15}{8}\frac{H}{\Gamma_d} \gg 1\,,
\end{equation}
which appears as a large prefactor in the DR Euler equation [Eq.~\eqref{eq:dr_euler}].

\subsection{Sub-horizon perturbation equations}

On sub-horizon scales ($k \gg \mathcal{H}$) during matter domination, with $\Psi = \Phi$, keeping only potential terms which are enhanced by $k^2$, and the background DR density attractor substituted, the perturbation equations simplify to
\begin{align}
k^2\Phi &= -\tfrac{3}{2}\mathcal{H}^2\!\left(r_\chi\,\delta_\chi + r_b\,\delta_b\right),\label{eq:app_poisson}\\
\dot\delta_\chi &= -\theta_\chi\,,\label{eq:app_dm_cont}\\
\dot\theta_\chi &= -\mathcal{H}\theta_\chi + k^2\Phi + a\Gamma_s\!\left(\theta_\mathrm{DR} - \theta_\chi\right),\label{eq:app_dm_euler}\\
\dot\delta_\mathrm{DR} &= -\tfrac{4}{3}\theta_\mathrm{DR} + \tfrac{5}{2}\mathcal{H}\!\left(\delta_\chi - \delta_\mathrm{DR}\right),\label{eq:app_dr_cont}\\
\dot\theta_\mathrm{DR} &= \tfrac{k^2}{4}\delta_\mathrm{DR} + k^2\Phi + \tfrac{15}{8}\mathcal{H}\!\left(\theta_\chi - \tfrac{4}{3}\theta_\mathrm{DR}\right)\nonumber\\
&\quad + \frac{15\,\Gamma_s}{8\,\Gamma_d}\,\mathcal{H}\!\left(\theta_\chi - \theta_\mathrm{DR}\right),\label{eq:app_dr_euler}
\end{align}
where $r_\chi = \Omega_\chi/\Omega_m$ and $r_b = \Omega_b/\Omega_m$.

\subsection{Perturbative solution for the DR}

The drag term in Eq.~\eqref{eq:app_dm_euler} couples the DM perturbations to the DR velocity $\theta_\mathrm{DR}$.
Since $\Gamma_s/H \ll 1$, the backreaction on DM is a small correction, and we can solve the DR equations to zeroth order in $\Gamma_s/H$ using the standard $\Lambda$CDM growing mode as input: $\delta_\chi^{(0)} \propto a$, $\theta_\chi^{(0)} \approx -\mathcal{H}\,\delta_\chi^{(0)}$, and $k^2\Phi^{(0)} \approx -\frac{3}{2}\mathcal{H}^2\,\delta_\chi^{(0)}$.
Substituting into the DR equations, expanding for large $k$ to order $\mathcal{H}^2/k^2$, and assuming $\Gamma_s/\Gamma_d\gg1$, the steady-state solution for the DR perturbations is
\begin{align}
\delta_\mathrm{DR} &\approx \frac{345\,\Gamma_s}{16\,\Gamma_d} \frac{\mathcal{H}^2}{k^2}\,\delta_\chi^{(0)}\,,\label{eq:app_dr_delta}\\
\theta_\mathrm{DR} &\approx \frac{15}{8}\,\mathcal{H}\,\delta_\chi^{(0)}\!\left(1 - \frac{345\,\Gamma_s}{16\,\Gamma_d}\frac{\mathcal{H}^2}{k^2}\right).\label{eq:app_dr_theta}
\end{align}
This solution shows that at short distances, $1/k\rightarrow 0$, the DR velocity divergence $\theta_\mathrm{DR}\approx 15/8\,\mathcal{H}\,\delta_\chi^{(0)}$ is positive in regions of DM overdensity. DR flows out of overdense regions due to its pressure, unlike DM which flows in due to gravitational attraction. Momentum transfer between the two fluids slows the growth of DM perturbations proportional to $\Gamma_s$. At larger distances $\frac{345\,\Gamma_s}{16\,\Gamma_d}\frac{\mathcal{H}^2}{k^2}\sim 1$, the outflow of DR reverses due to the coupling between the fluids. At very large scales $1/k \sim 1/\mathcal{H}$, the DR is strongly coupled to the DM, $\theta_\mathrm{DR} \rightarrow \theta_\chi$, and the drag on the DM velocity [Eq.~\eqref{eq:app_dm_euler}] disappears (the DR is dragged along by the DM, leaving no velocity difference to source a drag on the DM).

\subsection{Growth equation with scattering source}

The total cold matter perturbation $\delta_m = r_\chi\,\delta_\chi + r_b\,\delta_b$ satisfies the sourced growth equation
\begin{equation}
\label{eq:app_growth}
\ddot\delta_m + \mathcal{H}\dot\delta_m - \tfrac{3}{2}\mathcal{H}^2\,\delta_m = -r_\chi \cdot a\Gamma_s\!\left(\theta_\mathrm{DR} - \theta_\chi^{(0)}\right).
\end{equation}
The right-hand side is the drag-induced source, proportional to the difference in DM-DR fluid velocities. Substituting the approximate DR solution [Eq.~\eqref{eq:app_dr_theta}] and solving for the growing-mode solution during matter domination gives
\begin{equation}
\label{eq:app_deltam_ratio}
\frac{\delta_m}{\delta_m^{(0)}} \approx 1 - \frac{23}{48}\,r_\chi\,\frac{\Gamma_s}{H}\!\left(1 - \frac{225\,\Gamma_s}{4\,\Gamma_d} \frac{\mathcal{H}^2}{k^2}\right).
\end{equation}
This expression is valid for $k \gg k_\mathrm{d}$ and diverges as $k \to 0$.
Since we expect $\delta_m/\delta_m^{(0)} \to 1$ on large scales, we guess an Ansatz which interpolates between these limits
\begin{equation}
\label{eq:app_ansatz}
\frac{\delta_m}{\delta_m^{(0)}} \approx 1 - \frac{\mathcal{A}/2}{1 + (k_\mathrm{d}/k)^2}\,,
\end{equation}
with
\begin{align}
\mathcal{A} = \frac{23}{24}\,r_\chi\,\frac{\Gamma_s}{H}\,,\quad
k_\mathrm{d}^2 =\frac{\Gamma_s}{\Gamma_d}  \frac{15^2\,\mathcal{H}^2}{4} = \frac{\Gamma_s}{\Gamma_d}  \frac{15^2}{\tau^2},\label{eq:app_kstar_MD}
\end{align}
where the last step uses $\mathcal{H} = 2/\tau$ during matter domination. The expression for $k_\mathrm{d}$ in terms of conformal time also works reasonably well at $z=0$ even though our derivation assumed MD.

Squaring Eq.~\eqref{eq:app_ansatz} gives the power spectrum ratio
\begin{equation}
\label{eq:app_Pk_ratio}
\frac{P_m}{P_{\Lambda\mathrm{CDM}}} \approx 1 - \frac{\mathcal{A}}{1 + (k_\mathrm{d}/k)^2}\,.
\end{equation}

\begin{figure}[t]
\centering
\includegraphics[width=\columnwidth]{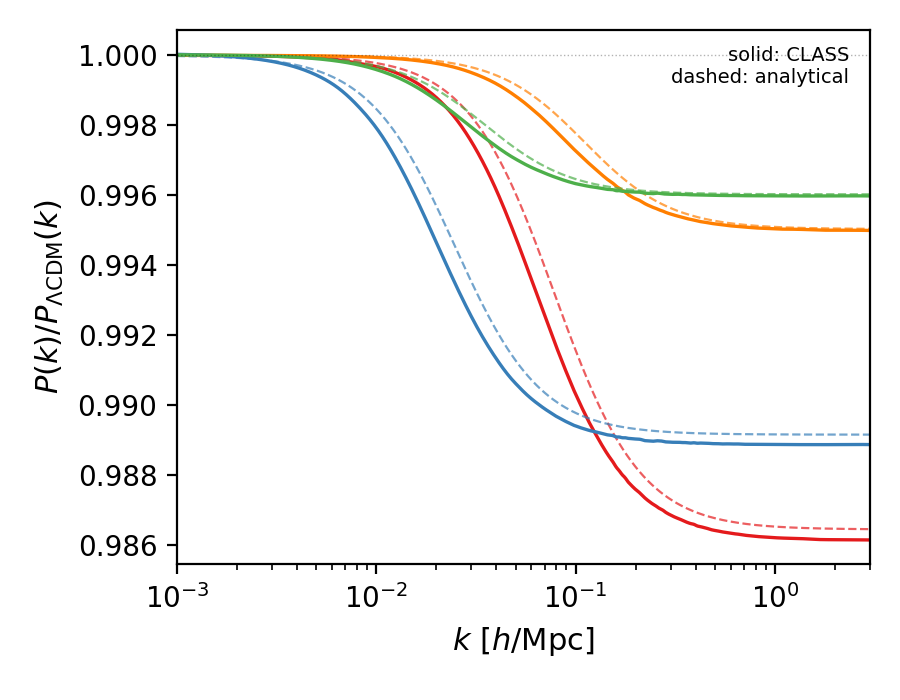}
\caption{Verification of the analytical approximation for $n_{\Gamma_s} = 0$.
Solid lines are \textsc{class}; dashed lines are Eq.~\eqref{eq:app_Pk_ratio}.
Curves with $\left(\Gamma_s^0/H_0,\Gamma_d/\Gamma_s^0\right)$ $= \left(0.05, 10^{-3}\right)$ are shown at $z=2$ (red) and $z=5$ (orange). Curves with $\left(\Gamma_s^0/H_0,\Gamma_d/\Gamma_s^0\right)$ $= \left(0.04, 10^{-2}\right)$ are shown at $z=2$ (blue) and $z=5$ (green).}
\label{fig:analytic_verify}
\end{figure}

Equation~\eqref{eq:app_Pk_ratio} was derived assuming MD and to first order in $\Gamma_s/H$ and $\Gamma_d/H$. It can be extended to late times where MD no longer holds. However, it is straightforward enough to solve the full perturbation equations and we do not pursue such an approximate solution here. Figure~\ref{fig:analytic_verify} shows that the analytical formula~\eqref{eq:app_Pk_ratio} captures the behavior of the full numerical result from \textsc{class} quite well as a function of $k$ for $ z \gtrsim 2$.

\section{UV Model Details and Rate Hierarchy}
\label{app:uv}

The UV model introduced in Sec.~\ref{sec:uv} contains the dark matter $\chi$, the massless fermion $\psi$, and the dark photon $\gamma_d$, interacting via the dark $U(1)$ gauge coupling $\alpha_d$.
Here we verify that the interaction rates satisfy the hierarchy required for the DR to behave as a perfect fluid and for the DM--DR coupling to be weak but become more important at low redshift.

\subsection{Light-particle thermalization}

The massless species $\psi$ and $\gamma_d$ self-interact with various 2 to 2 processes whose momentum transfer rates scale like
\begin{equation}
\label{eq:app_therm}
\Gamma_{\gamma\gamma} \sim \alpha_d^2\,T_d\,, \qquad \Gamma_{\psi\gamma} \sim \alpha_d^2\,T_d\,.
\end{equation}
Number-changing 2$\to$3 processes (e.g.\ double Compton) carry a naive extra $\alpha_d$ suppression, but an IR divergence from the massless fermion propagator, regulated by a thermal Debye mass, restores the rate to $\sim\alpha_d^2 T_d$.\footnote{Logarithmic factors $\log(1/\alpha_d)$ further enhance some estimates; we ignore them since they are not important.} Full kinetic and chemical equilibrium therefore requires $\alpha_d^2 T_d \gg H$, easily satisfied for any reasonable $\alpha_d$. The $\psi,\gamma_d$ plasma thermalizes to a relativistic perfect fluid with temperature $T_d$, energy density $\rho_\mathrm{DR}\propto T_d^4$, and vanishing anisotropic stress.

\subsection{DM--DR scattering}

The DM--DR momentum transfer rate from $\chi\,\psi \to \chi\,\psi$ scattering via dark photon exchange is [Eq.~\eqref{eq:Gammas_UV}]
\begin{equation}
\label{eq:app_Gammas}
\Gamma_s(T_d) \sim \alpha_d^2\,\frac{T_d^2}{M_\chi}\,,
\end{equation}
suppressed relative to DR-DR momentum transfer by a factor of $T_d/M_\chi\ll1$. Thus as long as the dark sector is cold compared with the mass of the dark matter, DR-DR interactions can easily be in equilibrium while DM-DR interactions can be weak. 

The $z$-dependence of $\Gamma_s$ traces the DR temperature. With DR in thermal equilibrium on the attractor, $T_d\sim\rho_\mathrm{DR}^{1/4}\sim(\rho_\chi\,\Gamma_d/H)^{1/4}\sim a^{-3/8}$, cooling much more slowly than decoupled radiation. In a two-stage decay extension ($\chi\to\chi'\to$ DR with $\Gamma'_d\sim\Gamma_d$), $\rho_\mathrm{DR}\sim\rho_\chi\,\Gamma'_d\Gamma_d/H^2\sim a^0$ and $T_d$ becomes constant in time. Motivated by this variation we parametrize
\bea
\Gamma_s(a)\equiv \Gamma_s^0 \left( \frac{\rho_\mathrm{DR}(a)}{\rho_\mathrm{DR}(1)}\right)^{n_{\Gamma_s}/4}
\eea
with $n_{\Gamma_s}=2$ corresponding to the $T_d^2$-dependence of the single-stage decay model and $n_{\Gamma_s}=0$ to the two-stage decay.

In order to gently suppress the growth of structure at $z = 0$ we require that $\Gamma_s^0/H_0 \sim \mathcal{O}(0.1)$. For example, for the $n_{\Gamma_s}=2$ model this implies
\begin{equation}
\label{eq:app_Gammas_H0}
\frac{\Gamma_s^0}{H_0} \sim \alpha_d^2\,\frac{T_{d,0}^2}{M_\chi\,H_0} \sim  \alpha_d^2\,\sqrt{\frac{\Gamma_d}{H_0}}\frac{M_{\rm Pl}}{M_\chi}\,,
\end{equation}
and choosing $\Gamma_s^0/H_0 \sim \sqrt{\Gamma_d/H_0} \sim 0.1$, we require $\alpha_d^2 \sim M_\chi/M_{\rm Pl}\ll 1$.

\subsection{Rate hierarchy}

Summarizing, we require the three rates to satisfy the ordering
\begin{equation}
\label{eq:app_hierarchy}
\alpha_d^2\,T_d \gg H \gtrsim \alpha_d^2\,\frac{T_d^2}{M_\chi}\,,
\end{equation}
which guarantees two essential properties.
First, the DR thermalizes on timescales much shorter than a Hubble time, validating the perfect-fluid treatment with $w = 1/3$ and $\sigma_\mathrm{DR} = 0$.
Second, the DM--DR momentum transfer rate is comparable to or smaller than $H$, so the elastic scattering produces a drag rather than tight coupling.
The hierarchy holds naturally for $M_\chi \gg T_d$, which is the regime where $\chi$ is non-relativistic dark matter.

\section{$P_m$ versus $P_\mathrm{cb}$}
\label{sec:pmpcb}

A subtlety arises when there are several kinds of matter with different cluster properties. For example, it is well-known that the massive $\Lambda$CDM neutrinos are non-relativistic today and their energy density scales as matter. Nonetheless neutrino velocities are well above escape velocities of even the largest structures so that massive neutrinos do not contribute to gravitational clustering. Thus neutrinos also do not contribute to the halos in which galaxies form. Therefore large scale structure observations which rely on observations of galaxy distributions are sensitive to the power spectrum of the clustering matter $P_\mathrm{cb}$~\cite{Castorina:2015bma}, without neutrinos.~\footnote{At the redshifts and $k$-scales of LSS observations baryons and CDM cluster equally and contribute to $P_\mathrm{cb}$.} On the other hand, observations which rely on gravitational lensing are sensitive to the full matter power spectrum including neutrinos $P_m$. In $\Lambda$CDM with neutrino masses $\sim 0.1$ eV this distinction is mostly academic because the energy density in the massive neutrinos is too small to produce a measurable step in the matter power spectrum and $P_m\simeq P_\mathrm{cb}$. 

However, the difference can be very important in models beyond $\Lambda$CDM with two kinds of DM with different clustering properties. Clustering dark matter contributes to both $P_m$ and $P_\mathrm{cb}$ while non-clustering matter only contributes to $P_m$. Whether a suppression in matter clustering due to non-clustering matter is observable depends on which power spectrum is being probed.
Galaxy clustering measurements and redshift-space distortions trace the power spectrum of clustering matter, $P_\mathrm{cb}(k)$.
Any matter species that does not cluster on the relevant scales contributes to $\Omega_m$ and enters $P_m$, but is invisible to galaxy surveys.

A natural alternative to the interacting DM-DR scenario discussed in this paper is two-body decay into a lighter dark matter species: $\chi_1 \to \chi_2 + \gamma_d$, where $\chi_2$ is a stable daughter that is still non-relativistic today. The daughter $\chi_2$ is produced with a large velocity set by the mass splitting, and assuming negligible scattering rates between any of the DM and DR, on sub-horizon scales it free-streams rather than clusters~\cite{Enqvist2015,Poulin1,Poulin2,Poulin3,Fuss:2022zyt,Fuss:2024dam}.
Because $\chi_2$ is matter, it contributes to $\Omega_m$ and enters $P_m(k)$; but since it does not cluster, it acts as a smooth component that dilutes $P_m$ relative to $\Lambda$CDM.
Galaxy surveys measure $P_\mathrm{cb}$, not $P_m$, so the suppression visible in $P_\mathrm{cb}$ is substantially smaller than in $P_m$~\cite{Poulin1}.
Weak lensing, by contrast, responds to the total matter distribution and therefore probes $P_m$ directly.
Such models predict two distinct power spectra, and by contrasting clustering or redshift-space distortions (which trace $P_\mathrm{cb}$) with lensing measurements (which trace $P_m$) one can discover the existence of non-clustering matter.

In iDCDM this subtlety does not arise.
The decay products ($\psi$ and $\gamma_d$) are massless; they do not contribute to $\Omega_m$ and do not enter $P_m$.
No non-clustering matter species is produced and $P_m(k) = P_\mathrm{cb}(k)$ (ignoring the small effects of massive neutrinos).
The suppression of the matter power spectrum comes from the drag on clustering matter [Eq.~\eqref{eq:dm_euler}] and is imprinted on $P_\mathrm{cb}$ directly.

\section{Posterior Distributions}
\label{app:posteriors}

\begin{figure}[H]
	\centering
	\includegraphics[width=0.85\textwidth]{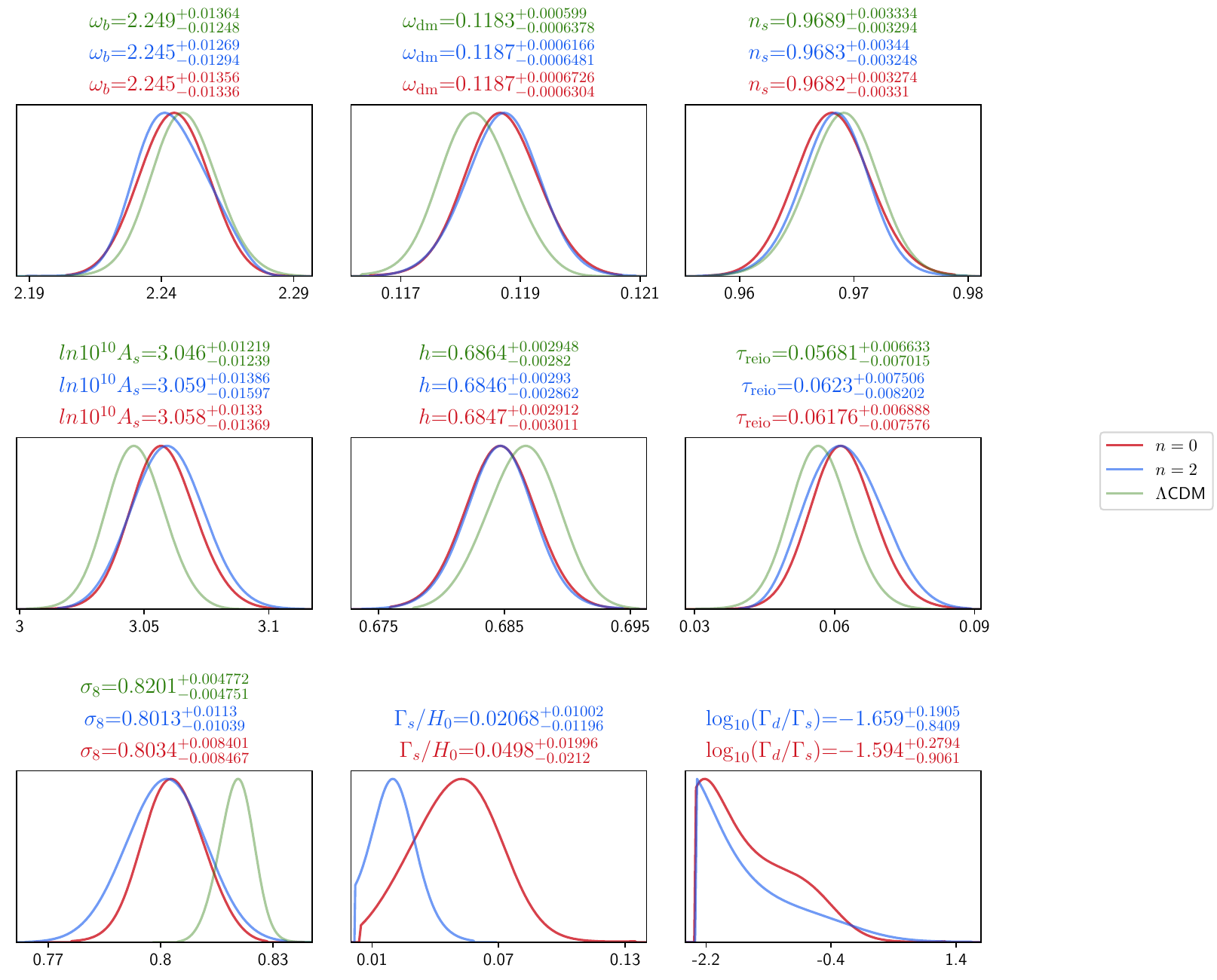}
	\caption{One-dimensional marginalized posterior distributions for $\Lambda$CDM (green), $n_{\Gamma_s} = 2$ (blue), and $n_{\Gamma_s} = 0$ (red), with massless neutrinos.}
	\label{fig:1d_posteriors_full}
\end{figure}

\begin{figure}[H]
\centering
\includegraphics[width=0.85\textwidth]{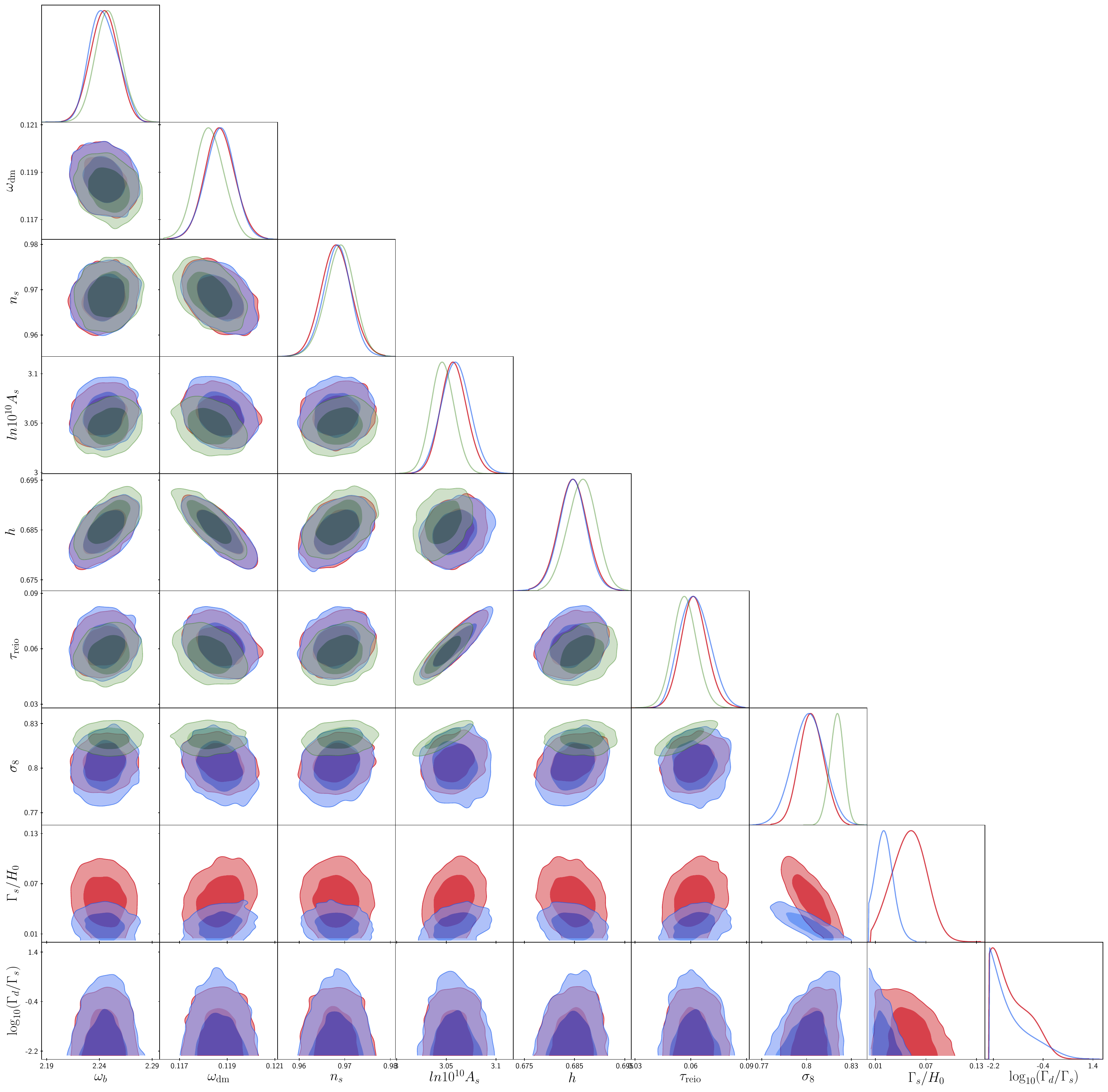}
\caption{Two-dimensional marginalized posterior contours (68\% and 95\% C.L.) and one-dimensional marginalized distributions for $\Lambda$CDM (green), $n_{\Gamma_s} = 2$ (blue), and $n_{\Gamma_s} = 0$ (red), with massless neutrinos.}
\label{fig:full_triangle}
\end{figure}

The one-dimensional marginalized posteriors and the full two-dimensional posterior contours for the massless-neutrino analysis are shown in Figs.~\ref{fig:1d_posteriors_full} and~\ref{fig:full_triangle}.
The interpretation of these distributions is discussed in Sec.~\ref{sec:results}.

\onecolumngrid
\clearpage
\clearpage
\twocolumngrid

\section{Statistical interpretation of the $\Delta\chi^2$ preference}
\label{app:stats}

The $\Delta\chi^2$ values quoted in the main text indicate the size of the iDCDM fit improvement over $\Lambda$CDM within the region of parameter space we trust. Translating $\Delta\chi^2$ into a frequentist significance or a Bayesian model-comparison statement requires assumptions that are not cleanly satisfied here.

\begin{figure}[t]
\centering
\includegraphics[width=\columnwidth]{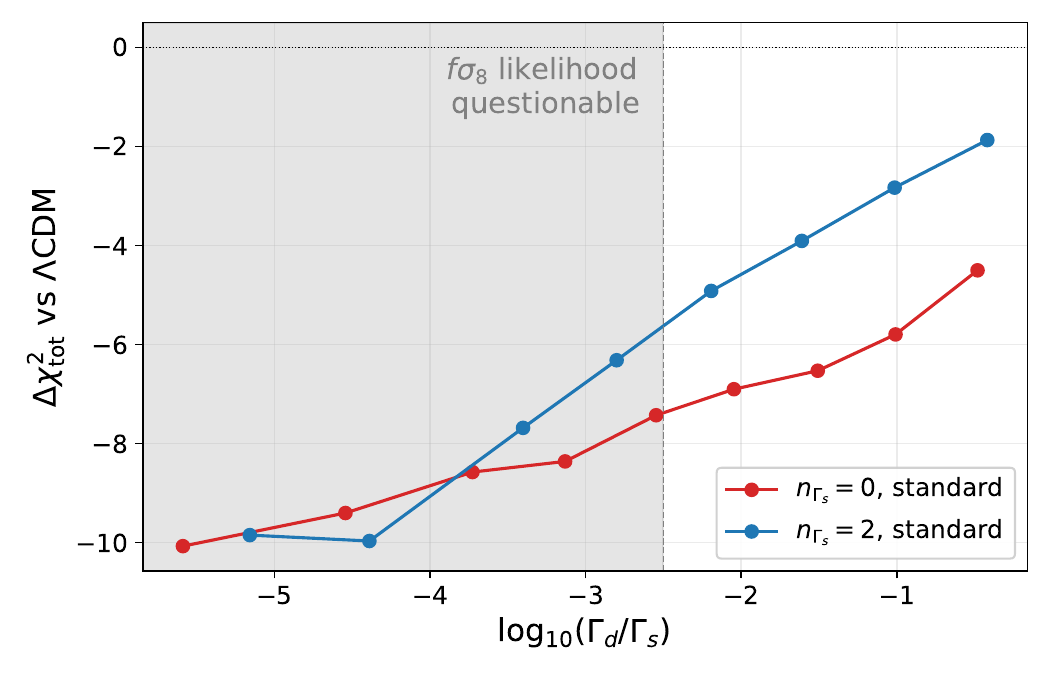}
\caption{Profile of the total $\Delta\chi^2$ relative to $\Lambda$CDM as a function of $\log_{10}(\Gamma_d/\Gamma_s)$, using the standard $f\sigma_8$ likelihood. Each grid point minimizes the full seven-likelihood $\chi^2$ over all other parameters at fixed $\log_{10}(\Gamma_d/\Gamma_s)$. The shaded region marks $\log_{10}(\Gamma_d/\Gamma_s) < -2.5$, where the standard $f\sigma_8$ likelihood becomes model-dependent (Fig.~\ref{fig:profile_fsig8}). Non-monotonicity at the $\Delta\chi^2 \lesssim 1$ level reflects finite convergence of the Powell minimizer at each grid point.}
\label{fig:profile_total}
\end{figure}

Figure~\ref{fig:profile_total} shows the profile of the total $\Delta\chi^2$ versus $\log_{10}(\Gamma_d/\Gamma_s)$. With the standard $f\sigma_8$ likelihood, the profile decreases past the cap at $-2.5$, indicating that the data would prefer a step at still larger $k_d$ if we allowed it. We hold the cap because past $-2.5$ the standard $f\sigma_8$ likelihood is no longer a faithful translation of $P(k)$ to the observable (Fig.~\ref{fig:profile_fsig8}); we leave the construction of a corrected likelihood to the survey teams with detailed knowledge of the data. The $\Delta\chi^2$ values in Table~\ref{tab:chi2_massless} should therefore be read as a conservative lower bound on the preference within the trusted region of parameter space, not as a detection significance.

The asymptotic identification of $\Delta\chi^2$ with a $\chi^2$ distribution requires the best fit to lie in the interior of the parameter space. In our analysis, $\log_{10}(\Gamma_d/\Gamma_s)$ sits at the prior boundary $-2.5$, which is not a physical boundary but a likelihood-validity edge: the available $f\sigma_8$ likelihood ceases to be a faithful description of the data past it (Fig.~\ref{fig:profile_fsig8}). The asymptotic calibration does not apply cleanly when the best fit is held against an edge of this kind.

We also do not report a Bayes factor. The marginal likelihood depends on the prior volume in the extra parameters; with our flat prior in $\log_{10}(\Gamma_d/\Gamma_s)$ truncated at $-2.5$, the result is set in part by a likelihood-validity cutoff rather than by a physically motivated prior, and the interpretation as model evidence is therefore not the standard one.

Finally, the diagonal $f\sigma_8$ errors used in the likelihood do not capture all sources of uncertainty. Independent analyses of overlapping low-redshift volumes return central values that differ by amounts comparable to their quoted statistical errors~\cite{Said_2020,Boruah_2020,Carrick_2015,Turner_2022}, and the analyses share survey volume so the errors cannot be independent at the level the diagonal-only treatment implies. Quantifying the resulting systematic floor would require a covariance model that is not provided with the published compilation. Together with the velocity-window issue of Fig.~\ref{fig:profile_fsig8}, this is an additional reason to treat the quoted $\Delta\chi^2$ as a qualitative indicator rather than a precise significance.

\bibliography{iDCDM}
\end{document}